\documentclass[osajnl,twocolumn,showpacs,superscriptaddress,10pt]{revtex4-1} 

\usepackage{dcolumn}
\usepackage{bm}
\usepackage{amsmath,amssymb,graphicx}
\usepackage[pagebackref=false,draft=false]{hyperref}
\usepackage{natbib}
\usepackage{gensymb} 
\usepackage{subcaption} 
\usepackage{color}

\newcommand{\titlestring}{Investigation of defect cavities formed in three-dimensional woodpile photonic crystals}

\newcommand{\EnergySnapshots}{frequency snapshots of the energy density}

\newcommand{\ElectromagneticEnergy}{$\varepsilon(\left|E_{x}\right|^{2}+\left|E_{y}\right|^{2}+\left|E_{z}\right|^{2})$}

\newcommand{\VerticalPeriod}{c}
\newcommand{\HorizontalPeriod}{a}

\newcommand{\RodWidth}{w}
\newcommand{\RodHeight}{h}

\newcommand{\SbXQ}{7.54\times10^{5}}
\newcommand{\SbXMVnormalized}{0.161 \cdot(\lambda_{res}/n)^{3}}

\newcommand{\AaXQ}{3.67\times10^{5}}
\newcommand{\AaXMVnormalized}{0.100 \cdot(\lambda_{res}/n)^{3}}

\newcommand{\ExampleLambdaNanometers}{ 638.98 }
\newcommand{\ExampleQfactor}{\SbXQ}

\newcommand{\VerticalPeriodOverLambda}{ 0.527 }
\newcommand{\RodWidthOverVerticalPeriod}{ 0.2145 }

\newcommand{\triono}{0.333} 

\newcommand{\KappaExprTwoLine}{
  $\begin{array}{c}
  \kappa_{uc}/(2\pi)\\
  =c_{0} / ( \lambda{}_{uc} \cdot{} Q_{uc} )
  \end{array}$ (GHz)
}

\newcommand{\KappaPrimeExprTwoLine}{
  $\begin{array}{c}
  \kappa{}'_{uc}/(2\pi)\\
  = c_{0} / ( \lambda{}_{os} \cdot Q_{uc} )
  \end{array}$ (GHz)
}

\newcommand{\eref}[1]{(\ref{#1})}
\newcommand{\sref}[1]{section~\ref{#1}}
\newcommand{\fref}[1]{figure~\ref{#1}}
\newcommand{\tref}[1]{table~\ref{#1}}

\newcommand{\Fref}[1]{Figure~\ref{#1}}
\newcommand{\Tref}[1]{Table~\ref{#1}}

\begin{document}

\hypersetup{pdftitle={\titlestring{}}}
\title{\titlestring{}}

\author{Mike P. C. Taverne}\email{Corresponding author: Mike.Taverne@bristol.ac.uk}
\author{Ying-Lung D. Ho}\email{Daniel.Ho@bristol.ac.uk}
\author{John G. Rarity}\email{John.Rarity@bristol.ac.uk}

\affiliation{Department of Electrical and Electronic Engineering, University of Bristol, University Walk, Bristol BS8 1TR, UK}

\begin{abstract}

We report the optimisation of optical properties of single defects in three-dimensional (3D) face-centred-cubic (FCC) woodpile photonic crystal (PC) cavities by using plane-wave expansion (PWE) and finite-difference time-domain (FDTD) methods.
By optimising the dimensions of a 3D woodpile PC, wide photonic band gaps (PBG) are created.
Optical cavities with resonances in the bandgap arise when point defects are introduced in the crystal.
Three types of single defects are investigated in high refractive index contrast (Gallium Phosphide-Air) woodpile structures and Q-factors and mode volumes ($V_{eff}$)  of the resonant cavity modes are calculated.
We show that, by introducing an air buffer around a single defect, smaller mode volumes can be obtained.
We demonstrate high Q-factors up to 700000 and cavity volumes down to $V_{eff}<0.2(\lambda/n)^3$.
The estimates of $Q$ and $V_{eff}$ are then used to quantify the enhancement of spontaneous emission and the possibility of achieving strong coupling with nitrogen-vacancy (NV) colour centres in diamond.

\end{abstract}


\ocis{(160.5298) Photonic crystals; (140.3945) Microcavities; (270.5580) Quantum electrodynamics; (350.4238) Nanophotonics and photonic crystals.}

\maketitle{} 

\section{Introduction}

In recent years, there has been remarkable progress in designing and fabricating three-dimensional (3D) photonic crystal based optical micro-cavities containing suitable emitters\cite{Ho2011:IEEE_JQE,Tandaechanurat2010,Aoki2003,Ishizaki2009,Qi2004,Lin2005,Braun2006,Ozbay1994}.
Such systems show long photon life times (i.e. high Q-factors) and small mode volumes ($V_{eff}$).
The resulting strength of interaction between the cavity photons and a two-level system leads to dramatic modifications of spontaneous emission \cite{Purcell1946,Benisty1998a,Hennessy2007,Lodahl2004} and cavity quantum electro dynamical (cavity-QED) phenomena \cite{Thompson1992,Reithmaier2004a,Peter2005a,Hennessy2007,Englund2007} known generically as strong coupling.
Potential applications of these devices include electro-optic modulators \cite{Tanabe2009, Zhang2013}, photonic sensors \cite{Chakravarty2014, Zhang2014}, ultra-small optical filters \cite{Song2003}, nonlinear optical devices \cite{Corcoran2009}, ultralow-power and ultrafast optical switches \cite{Nozaki2010}, quantum information processing \cite{Xiao2008,Young2009,Young2011,Hu2011} and low threshold lasers\cite{Loncar2002}.
Some of these achievements have been reported using two-dimensional (2D) PC slab cavities \cite{Fujita2005} and waveguides \cite{Schwagmann2011}, which are easier to fabricate and simulate.
However, in 2D PC designs, it is difficult to design low volume high finesse cavities due to the out-of plane radiation losses and scattering.
In contrast, 3D PC designs provide stronger confinement due to their complete photonic band gap \cite{Yablonovitch1987,Ho1990} leading to smaller mode volumes thus larger effects with weaker emitters.

Recently, several 3D PC structures with complete PBGs have been simulated and fabricated using two-photon polymerization (2PP) based 3D lithography or direct laser writing (DLW) system, with or without backfilling materials to create 3D high-refractive-index-contrast nanostructures \cite{Deubel2004:nanoscribe,VonFreymann2010,website:nanoscribe}.
In this work, we investigate the potential to create high-Q low mode volume cavities using the optical properties of single defects in three-dimensional (3D) face-centred-cubic woodpile\cite{Ho1994:OriginalWoodpile} photonic crystal cavities.
Analysis using the plane-wave-expansion (PWE) method has shown a full photonic bandgap in technically interesting wavelength regions.
We have studied the optimisation of relative gap width (gap width to midgap frequency ratio) as a function of the rod width using MIT Photonic-Bands software (MPB) \cite{Johnson2001:mpb,website:MPB}.
Our in-house finite-difference time-domain software \cite{Railton:BristolFDTD} is then used to calculate the time response and transmission spectrum of the finite woodpile crystal structures containing three types of single defect cavities with and without an air buffer to further isolate the cavity.
We demonstrate high Q-factors up to 700000 - limited only by simulation volume constraints - and cavity volumes down to $V_{eff}<0.2(\lambda/n)^3$.
Our Q-factors are comparable to other reported structures \cite{McCutcheon2008,Vahala2003,vahala-ref-48,vahala-ref-62}, but we note that it is the much smaller cavity volumes, which are of interest here.
These cavity volumes are an order of magnitude smaller than recent woodpile cavity simulations \cite{Woldering2014} and a factor of four smaller than nanobeam cavities \cite{McCutcheon2008}.
By introducing an air buffer around the defect, even smaller mode volumes can be obtained with slightly lower Q-factors.
We then estimate the enhancement of spontaneous emission and possibility of achieving strong coupling.
As an example system, we use emission rates estimated from nitrogen-vacancy (NV) colour centres in nanodiamond embedded in 3D FCC woodpile PhC structures made of gallium phosphide.


\section{Geometry and bandgaps: 3D FCC woodpile photonic crystal cavity design}

\subsection{Geometry: Design of the model \label{section:geometry}}

\begin{figure}
  \centering

  \begin{subfigure}[b]{0.5\linewidth}
    \centering
    \includegraphics[width=0.99\textwidth]{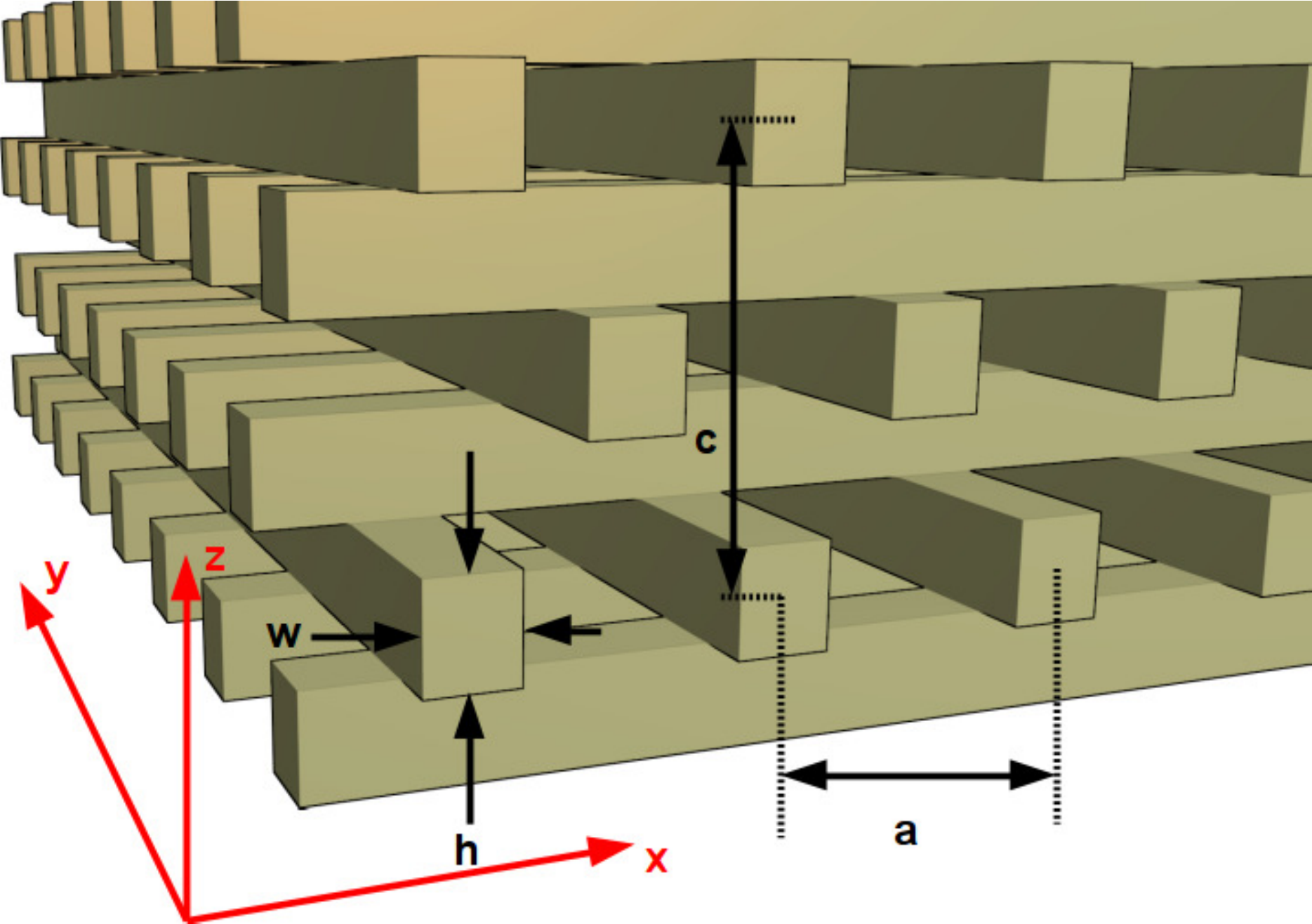}
    \caption{}
    \label{fig:geometry-parameters}
  \end{subfigure}%
  \begin{subfigure}[b]{0.5\linewidth}
    \centering
    \includegraphics[width=0.99\textwidth]{figure01b}
    \caption{}
    \label{fig:geometry-defect}
  \end{subfigure}%

  \caption{(a) Illustration of the woodpile geometry and its parameters. (b) Illustration of the defect and air buffer. The defect is shown in red with dimensions $(d_{x},d_{y},d_{z})$ and the boundaries of the air buffer in black with dimensions $(b_{x},b_{y},b_{z})$.}

\end{figure}

We are studying the woodpile structure as shown in \fref{fig:geometry-parameters}.
A woodpile structure consists of layers of parallel dielectric rods, with each layer rotated by 90\degree{} relative to the other.
Additionally, the layers are shifted by half a period every 2 layers.
This means that the structure repeats itself in the stacking direction every 4 layers.
We define the period along the stacking direction, corresponding to 4 layers, as $\VerticalPeriod{}$ and the distance between two rods within each layer as $\HorizontalPeriod{}$.
In our case, the rods have a rectangular cross-section with a height $\RodHeight{}$ and width $\RodWidth{}$. For a non-infinite woodpile crystal, we also define the number of layers $N_{l}$ and the number of rods per layer $N_{r/l}$.
In our simulations, to make the woodpile more symmetric, the number of rods per layer actually varies from $N_{r/l}$ to $N_{r/l}+1$. 

\begin{figure}

  \begin{subfigure}[b]{0.5\linewidth}
    \includegraphics[width=0.99\linewidth]{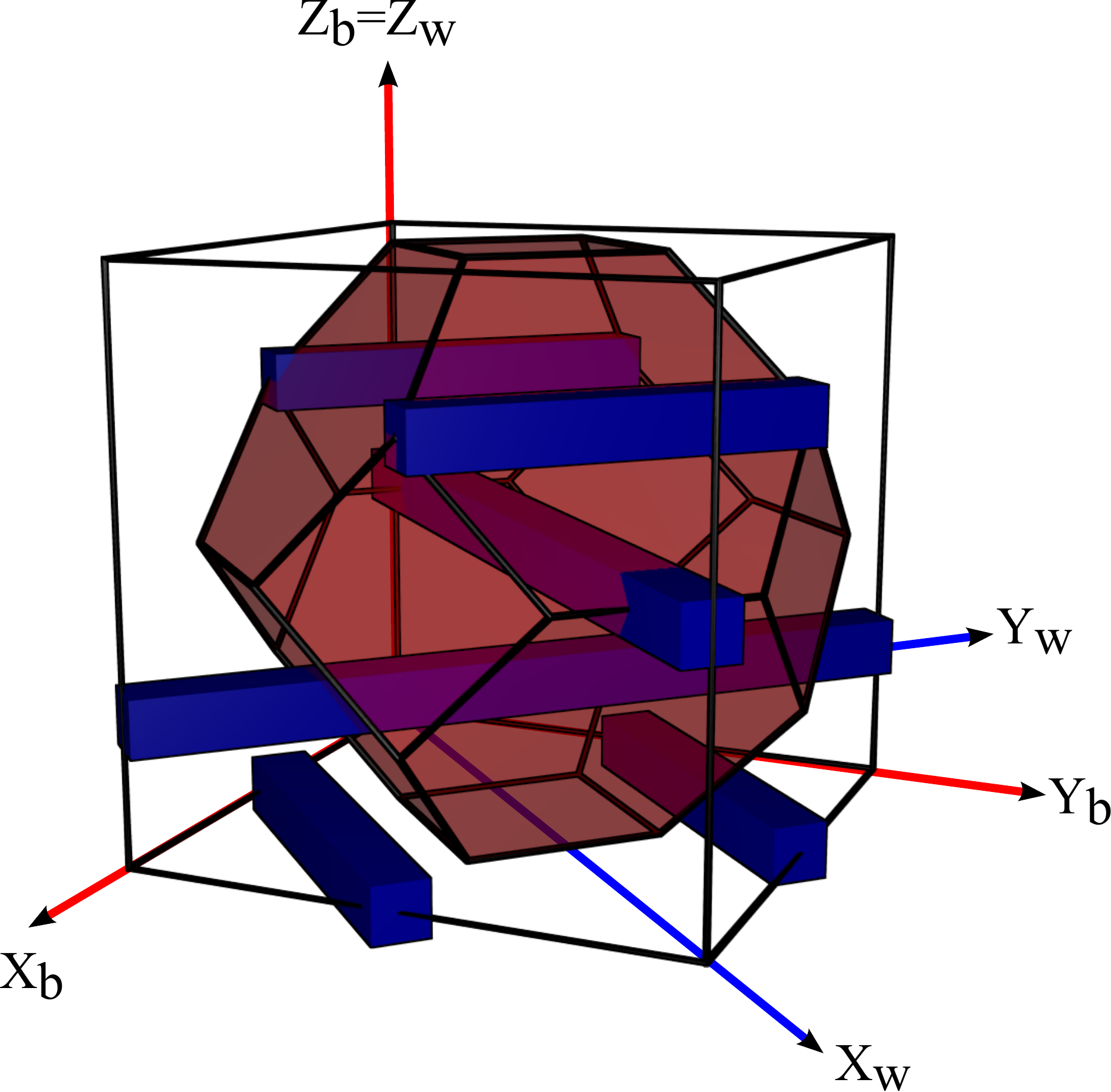}
    \caption{}
    \label{fig:WPinBZ}
  \end{subfigure}%
  \begin{subfigure}[b]{0.5\linewidth}
    \includegraphics[width=0.99\linewidth]{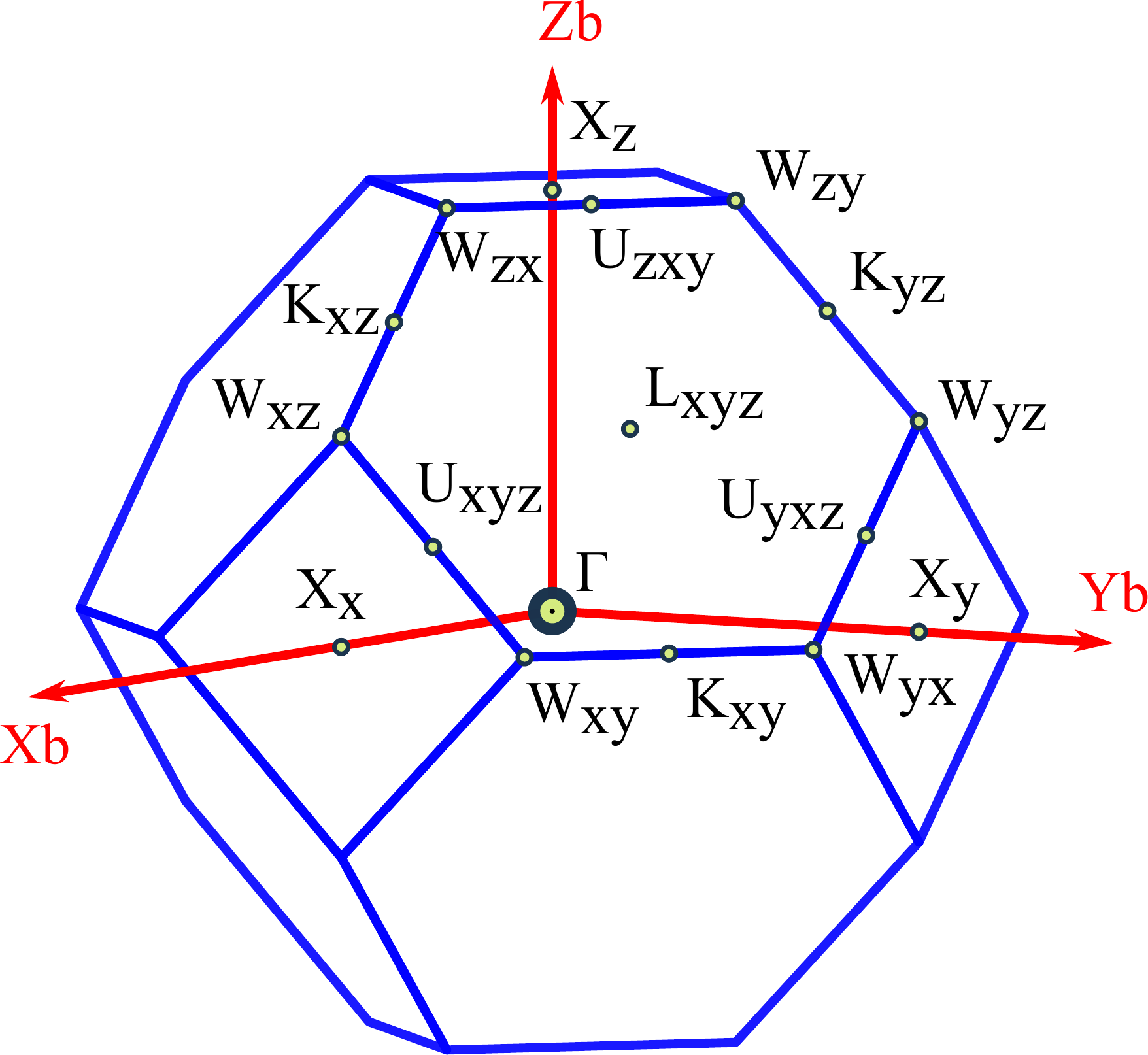}
    \caption{}
    \label{fig:BZlabels}
  \end{subfigure}%

  \caption{(a) The orientation of the woodpile relative to the FCC brillouin zone. The size of the rods has been changed to improve visibility. (b) Labeling of the traditional FCC brillouin zone points $\Gamma{}$, X, U, L, K and W relative to fixed brillouin zone axes $X_{b},Y_{b},Z_{b}$.}

\end{figure}

Depending on the ratio $\VerticalPeriod{}/\HorizontalPeriod{}$, the woodpile corresponds to different crystal lattices.
For $\VerticalPeriod{}/\HorizontalPeriod{} = 1$, it corresponds to a body-centered-cubic (BCC) lattice, for $c/a=\sqrt{2}$, to a face-centered-cubic (FCC) lattice and otherwise to a centered-tetragonal lattice. Here we chose $c/a=\sqrt{2}$, i.e. an FCC lattice as this has the most 'spherical' Brilluoin zone and thus produces the widest bandgaps.
\Fref{fig:WPinBZ} illustrates the Brillouin zone of the FCC lattice relative to the woodpile.
While the FCC woodpile shares the translational symmetry of an FCC lattice, it does not have the same rotational symmetries. The usual ``X points'' of the FCC brillouin zone are no longer equivalent.
We therefore used more specific labels for the FCC brillouin zone as detailed in \fref{fig:BZlabels}.
We orient the brillouin zone in the $(X_{b},Y_{b},Z_{b})$ basis and the woodpile in the $(X_{w},Y_{w},Z_{w})$ basis with $X_w = (X_b+Y_b)/\sqrt{2}$, $Y_w = (-X_b+Y_b)/\sqrt{2}$, $Z_w=Z_b$.
The layers in our woodpiles are stacked along the $Z_{w}$ direction and the rods aligned along the $X_{w}$ and $Y_{w}$ directions.

In order to create a resonant cavity in which the photons will be confined, a defect has to be added to the crystal \cite{Ho2011:IEEE_JQE,Ishizaki2013,Woldering2014}.
Following our previous work \cite{Ho2011:IEEE_JQE}, adding sphere based defects to air sphere crystals, here we choose a cuboid defect.
The defect position is shown in \fref{fig:geometry-defect}.
It is centered between two rods of the middle layer (with rods along the $X_{w}$ axis) and positioned along $X_{w}$ so that it is situated under a rod from the layer directly above it.

\begin{figure}
  \centering

  \begin{subfigure}[b]{\triono\linewidth}
    \centering
    \includegraphics[width=0.99\textwidth]{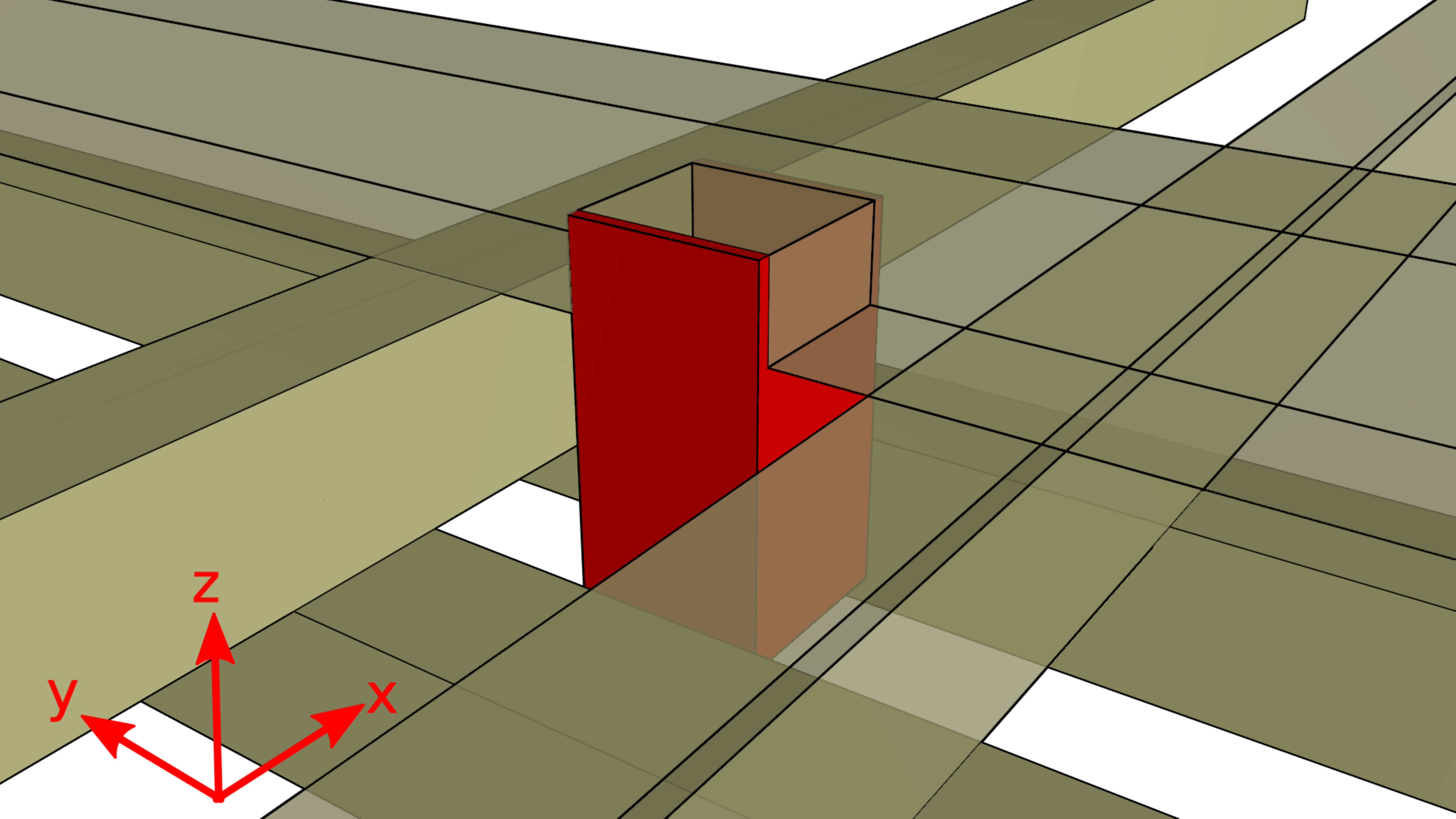}
    \caption{D0}
    \label{fig:D0}
  \end{subfigure}%
  \begin{subfigure}[b]{\triono\linewidth}
    \centering
    \includegraphics[width=0.99\textwidth]{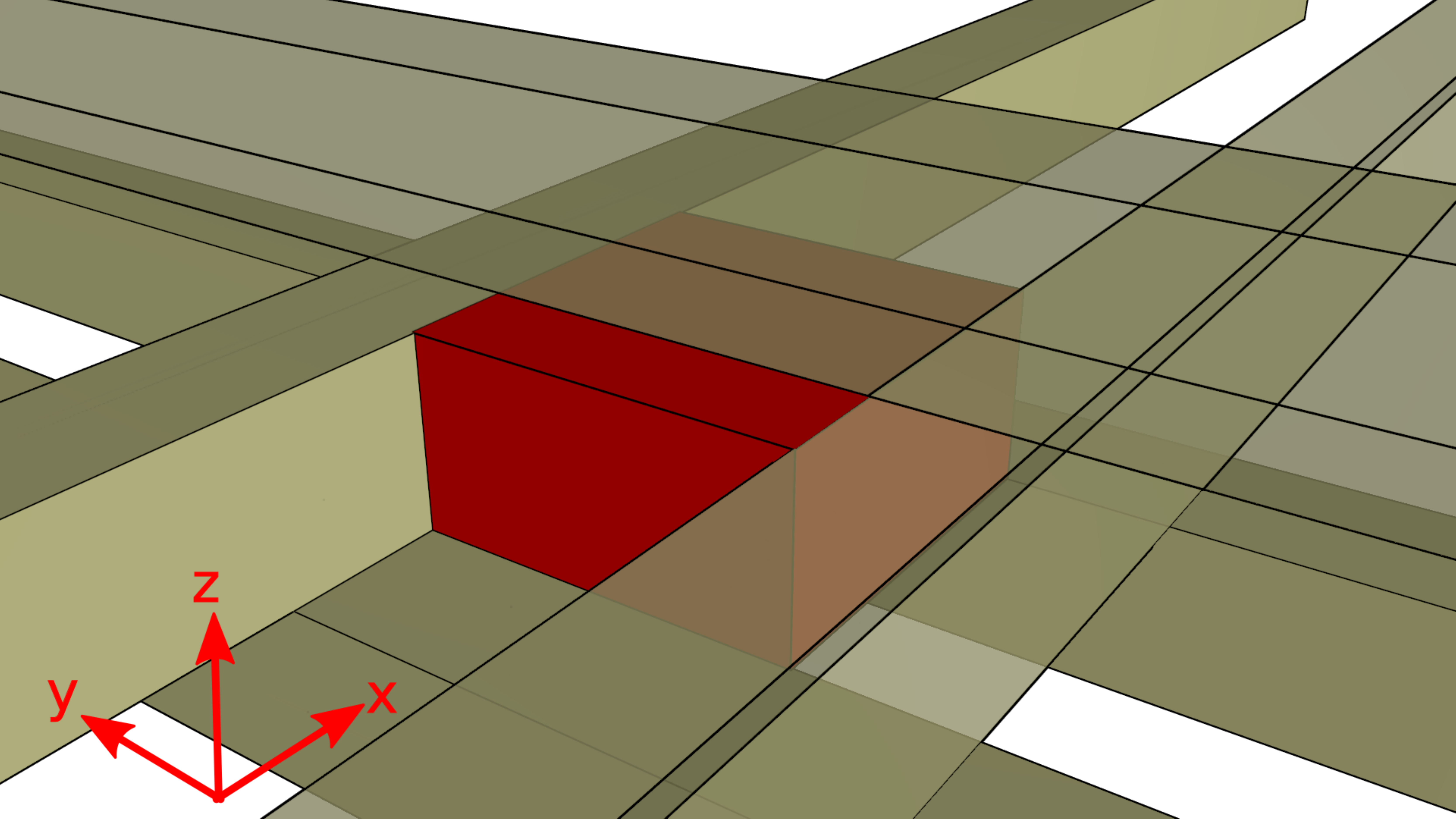}
    \caption{D1}
    \label{fig:D1}
  \end{subfigure}%
  \begin{subfigure}[b]{\triono\linewidth}
    \centering
    \includegraphics[width=0.99\textwidth]{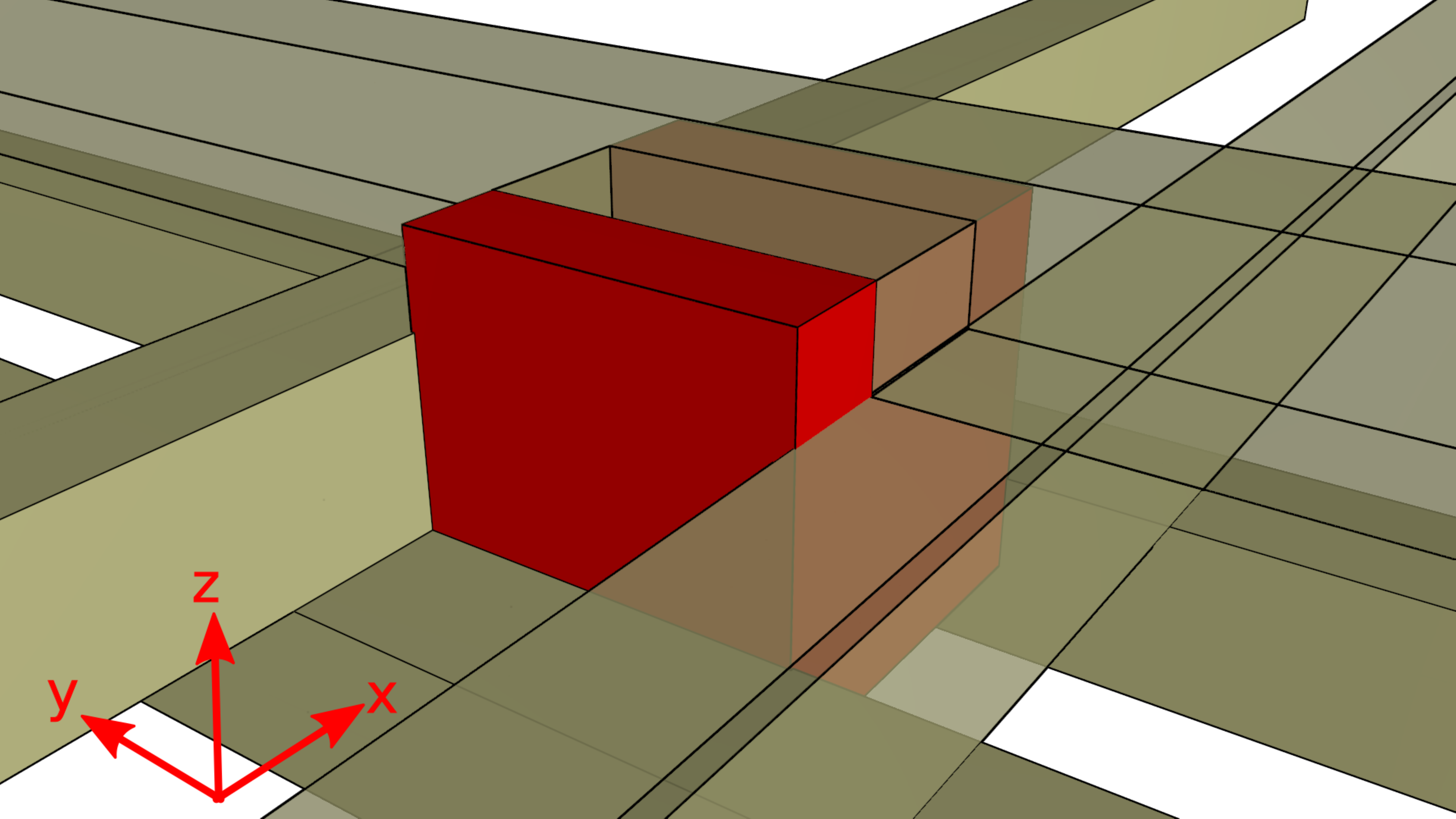}
    \caption{D2}
    \label{fig:D2}
  \end{subfigure}%

  \caption{The defects without air buffer D0 (a), D1 (b), D2 (c), of sizes $(d_{x},d_{y},d_{z}) = (0.25,0.25,0.5) \cdot c$, $(0.5,0.5,0.25) \cdot c$ and $(0.5,0.5,0.5) \cdot c$ respectively.}

\end{figure}

\begin{figure}

  \begin{subfigure}[b]{\triono\linewidth}
    \centering
    \includegraphics[width=0.99\textwidth]{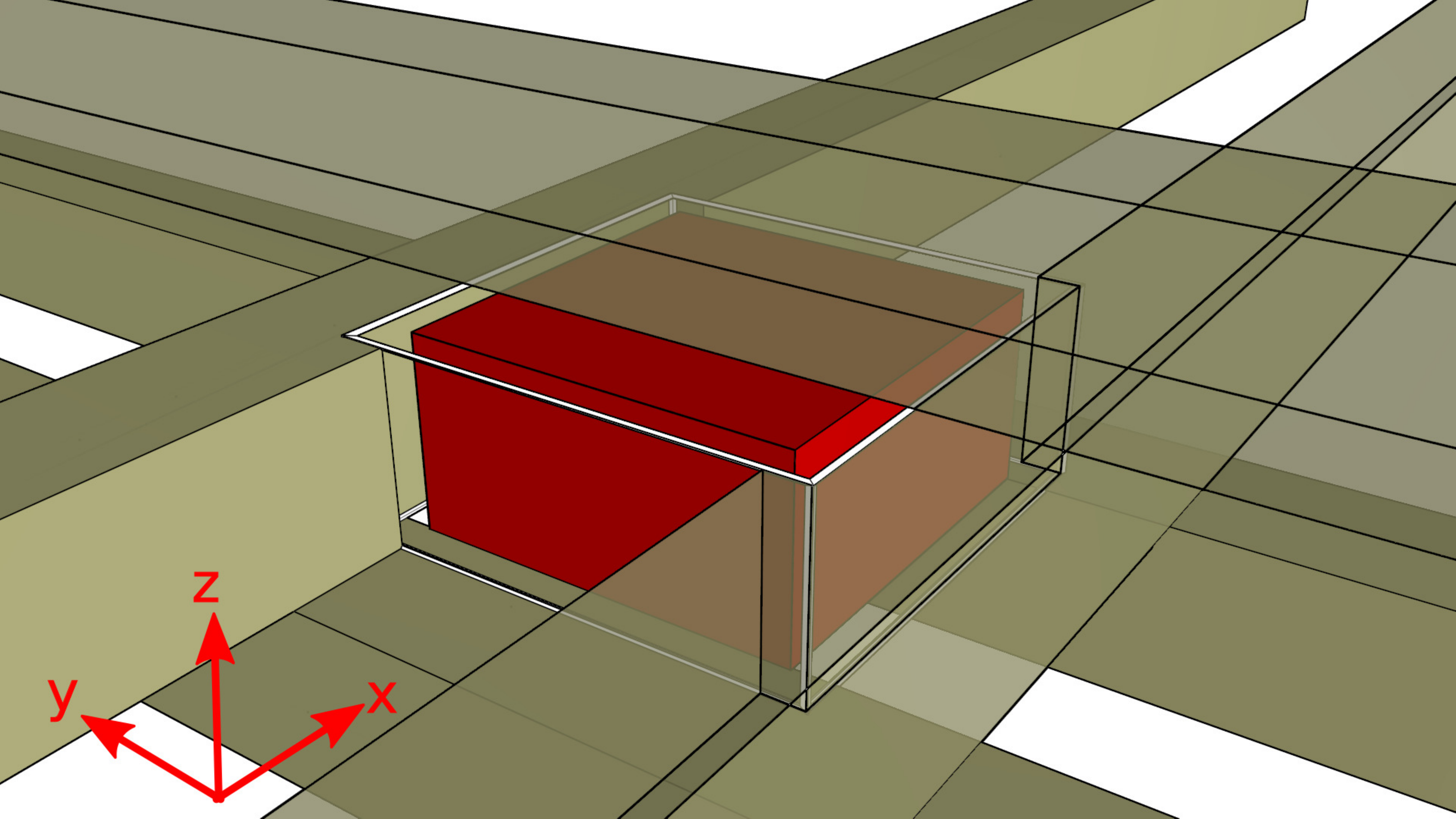}
    \caption{A0}
    \label{fig:A0}
  \end{subfigure}%
  \begin{subfigure}[b]{\triono\linewidth}
    \centering
     \includegraphics[width=0.99\textwidth]{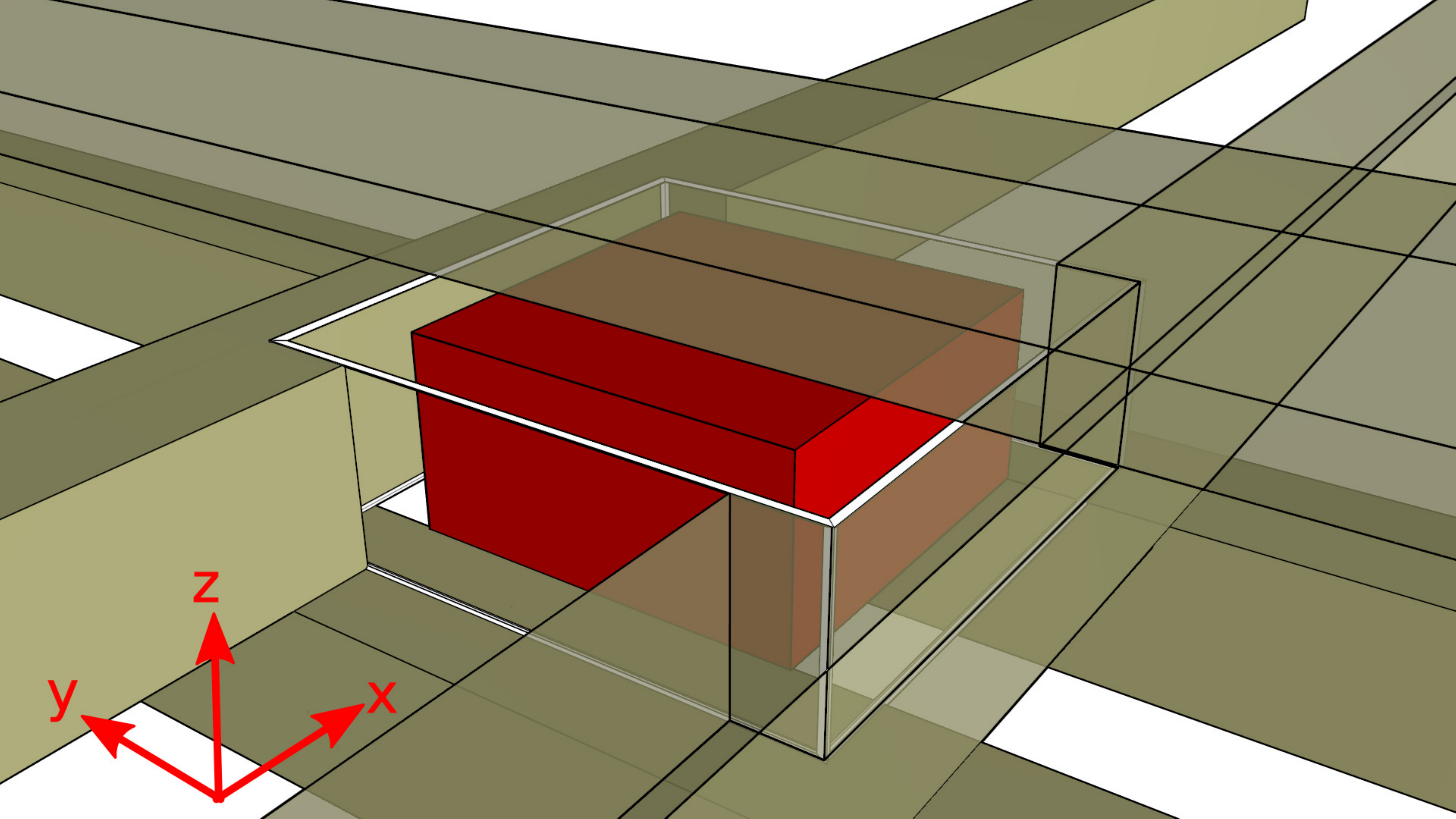}
    \caption{A1}
    \label{fig:A1}
  \end{subfigure}%
  \begin{subfigure}[b]{\triono\linewidth}
    \centering
    \includegraphics[width=0.99\textwidth]{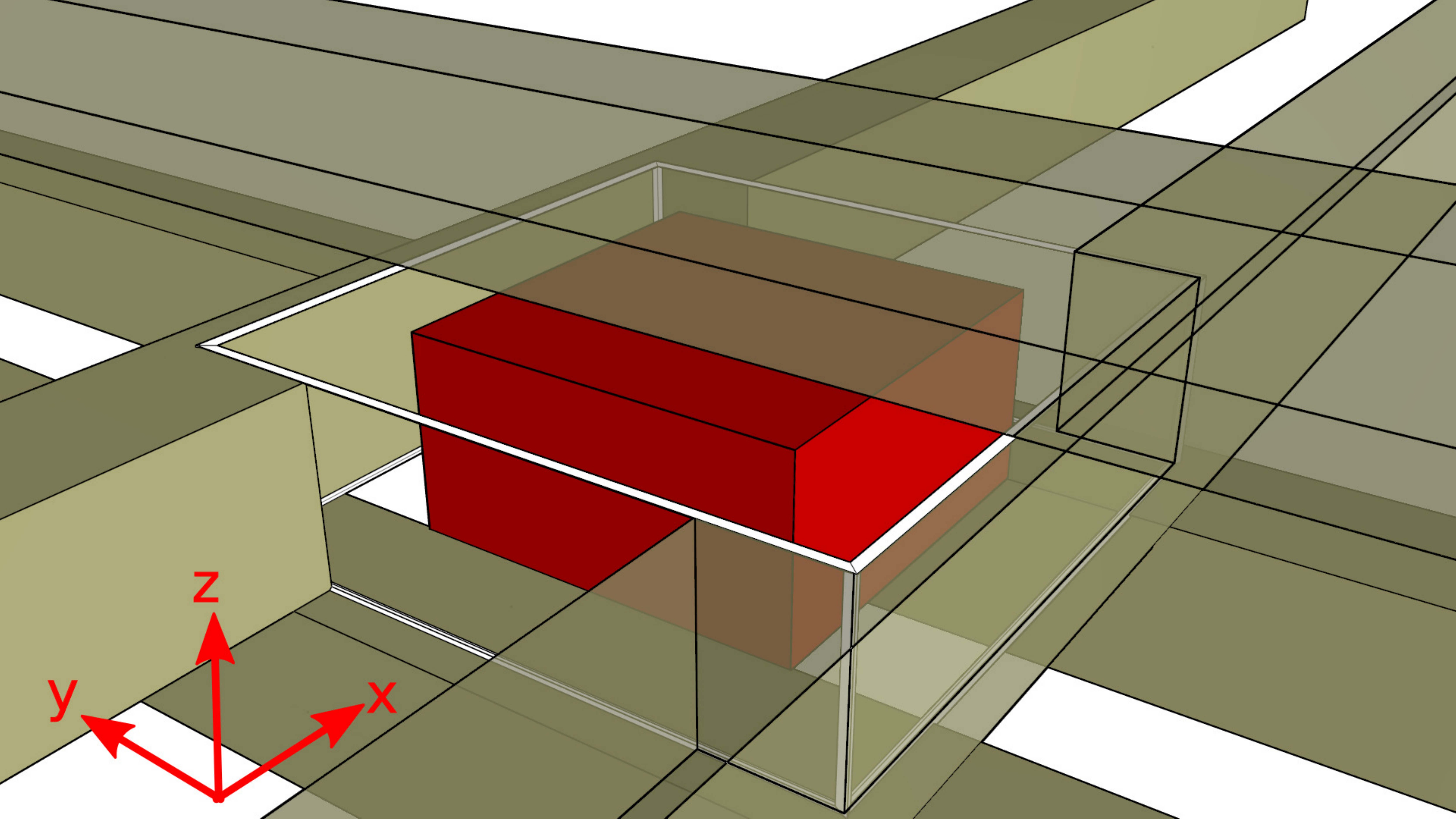}
    \caption{A2}
    \label{fig:A2}
  \end{subfigure}%

  \caption{The defect D1 with air buffers A0 (a), A1 (b), A2 (c) of sizes $(b_{x},b_{y},b_{z}) =$ ($a - 0.5 \cdot w$, $a - 0.5 \cdot w$, $0.25 \cdot c$), $(a, a, 0.25 \cdot c)$ and $(a + 0.5 \cdot w, a + 0.5 \cdot w, 0.25 \cdot c)$ respectively. The boundaries of the air buffers are indicated by the white wireframe box.}

\end{figure}

We initially conducted simulations for a fixed defect size, but at different positions within the woodpile.
First, so that it is not directly over/under a rod from the adjacent layer, second with the defect placed on a rod.
We found that the confinement properties were worse (lower Q-factor, larger mode volume) than for the position between the rods (described above and in \fref{fig:geometry-defect}) thus we chose this symmetry for all the work reported here.
However, we consider three different defect sizes D0, D1 and D2, as illustrated in figures \ref{fig:D0}, \ref{fig:D1} and \ref{fig:D2}.
In an attempt to reduce the energy leakage from the defect into the woodpile, we also looked at the effect of adding a cuboid air buffer around the defect D1 (i.e. in the same position, but larger than the defect).
Three different air buffer sizes A0, A1 and A2, as illustrated in figures \ref{fig:A0}, \ref{fig:A1} and \ref{fig:A2} were considered.
For our simulations, we used $\RodHeight{}/\VerticalPeriod{}=1/4$ and an experimentally relevant refractive index $n=n_{wp}=n_{def}=3.3$ (corresponding to Gallium-Phosphide (GaP), for example) for the woodpile rods and the defect.
The air buffer and the backfill material is considered to be air/vacuum of refractive index $n_{bf}=1$.
For the non-infinite woodpile crystal, we used $N_{l}=37$ and $N_{r/l}=13$.

\subsection{Bandgaps: Photonic bands of a 3D FCC woodpile PhC without defects}

\begin{figure*}
  \centering

  
  \begin{subfigure}[t]{0.5\linewidth}
    \includegraphics[width=0.99\textwidth]{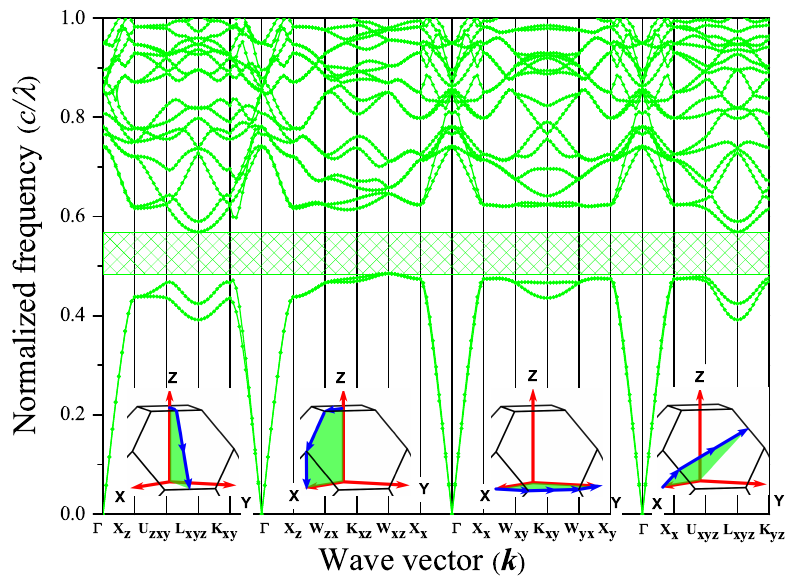}
    \caption{}
    \label{fig:bandgap-diagram}
  \end{subfigure}%
  \begin{subfigure}[t]{0.5\linewidth}
    \includegraphics[width=0.99\textwidth]{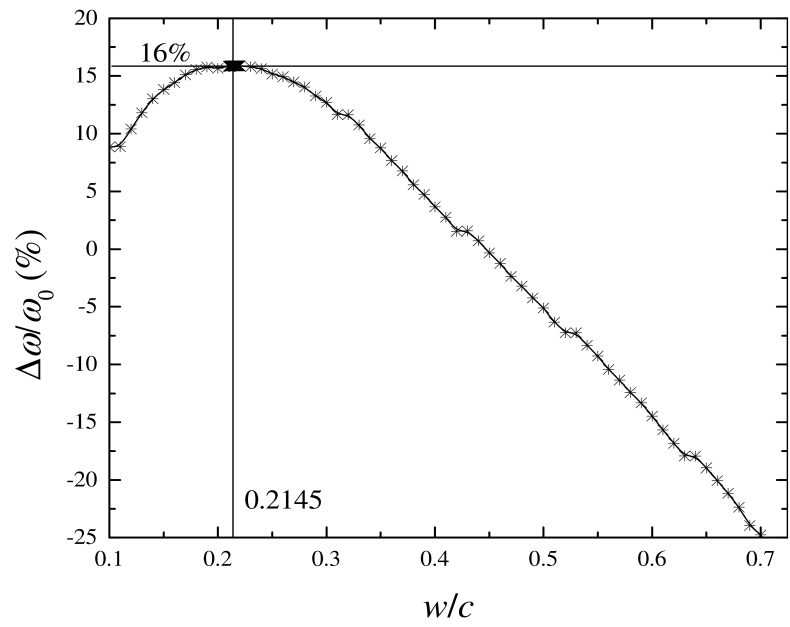}
    \caption{}
    \label{fig:bandgap-vs-w}
  \end{subfigure}%

  \caption{
    (a) Bandgap diagram of a woodpile with rods of refractive index 3.3 and a backfill of refractive index 1 (vacuum). Created with the MPB software \cite{Johnson2001:mpb,website:MPB}. The labels used for the different wavevectors $\vec{k}$ are explained in \fref{fig:BZlabels}.
    (b) Gap-midgap ratio $\Delta \omega / \omega_{0}$ (between band 2 and band 3) as a function of $\RodWidth{}/\VerticalPeriod{}$, where $\RodWidth{}$ is the width of the woodpile rods and $\VerticalPeriod{}$ its ``stacking period'', as explained in \fref{fig:geometry-parameters}.
  }
\end{figure*}

Initially, it is necessary to find the allowed propagation modes and eventual bandgaps of our photonic crystal. To do this, we used the MIT photonic bands package, which solves the frequency domain eigenproblem for a periodic dielectric structure \cite{Johnson2001:mpb,website:MPB}, giving us the allowed propagation modes for different wavevectors $\vec{k}$. The results for the woodpile structure described above using $(\RodWidth{}/\VerticalPeriod{})_{opt} = 0.2145$ are shown in \fref{fig:bandgap-diagram}. A full bandgap from $ \VerticalPeriod /\lambda_{0} \simeq 0.4853$ to $0.5689$ is clearly seen.

This value of $(\RodWidth{}/\VerticalPeriod{})_{opt}$ was chosen after calculating the gap-midgap ratio $\Delta \omega / \omega_{0}$ as a function of $\RodWidth{}/\VerticalPeriod{}$.
\Fref{fig:bandgap-vs-w} shows the results of these calculations, with a maximum  gap-midgap ratio of $~16\%$ for $(\RodWidth{}/\VerticalPeriod{})_{opt}$.

\section{Calculating relevant parameters of the D1 cavity}

In the following, we show how we calculate relevant parameters for our cavity structures.
These are the cavity resonant wavelength $\lambda_{0}$, the quality factor $Q$, the mode volume $V_{eff}$, the Purcell factor  $F_{p}$ and the coupling strength $g_{R}$.
We use as detailed example the D1 cavity described earlier in \fref{fig:D1}.

\subsection{Resonant wavelength and Q-factors of cuboid defect cavities with and without air buffers}

Now that the frequency range of the full photonic bandgap has been ascertained, we will look at the confinement efficiency of different defect cavities within this woodpile.
To do this we use a finite woodpile with a defect, as previously described in \sref{section:geometry}.
We then add a broadband source covering the bandgap within the defect and look at the time decay of the electromagnetic field using FDTD.
In order to better distinguish modes, we simulate using three orthogonal orientations of the dipole in 3D space: along the $X_{w}$, $Y_{w}$ and $Z_{w}$ axis, which we designate by $E_{x}$, $E_{y}$ and $E_{z}$ respectively. 

The centre of the bandgap is at $ \VerticalPeriod / \lambda_{0} \simeq \VerticalPeriodOverLambda $.
For $ \lambda_{0} = 637nm $, which corresponds to the zero-phonon line (ZPL) of nitrogen-vacancy centres (NV centres) in diamond, the corresponding vertical period of the woodpile is $ \VerticalPeriod = \VerticalPeriodOverLambda \cdot\lambda_{0} = 336 nm$.
The other parameters are then $ \HorizontalPeriod = \VerticalPeriod/\sqrt{2} \simeq 237nm $ for the horizontal period inside each layer, $ \RodWidth = \RodWidthOverVerticalPeriod \cdot \VerticalPeriod \simeq 72nm$ for the rod width and $ \RodHeight = 0.25 \cdot \VerticalPeriod \simeq 84nm $ for the rod height.

In the simulation a field probe is placed close to the excitation dipole located $1/4 d_{x}$, $1/4 d_{y}$ or $-1/4 d_{z}$ away from it depending on its direction.
The field at this probe is monitored as the simulation progresses and we obtain a slowly decaying field amplitude oscillating primarily at the allowed frequencies of the cavity.
Hence taking the fast Fourier transform (FFT) of this ringdown signal allows us to determine the resonant frequencies of the cavities and gives us an estimate of the Q-factors of these resonances ($Q = \lambda / \Delta \lambda $).

As these are high-Q (long time decay) resonances, the FDTD simulations were run for $ \sim 10^{5}-10^{6} $ iterations (timestep $ \delta t \sim 3-11 \times 10^{-18} s $, total time $ \sim 1-7 \times 10^{-12} s $) in order to confirm the field decayed significantly by the end of the simulation.
A single simulation using a non-homogeneous mesh of around $10^{7}$ cells adapted to the geometry takes about 2 weeks on a computing node with two 2.8 GHz quad-core Intel Harpertown E5462 processors, using a single core (as unfortunately, our software does not support parallel processing) and 2 out of the available 8GB of RAM memory\cite{website:ACRC}. The field decay and spectrum for a woodpile with defect D1 is shown in \fref{fig:probeSignal} illustrating the simple method of calculating Q. In many structures, we do not achieve a full decay of the field: this limits the Q values measured from the FFT peaks. Hence, for the analysis of very long lived decays corresponding to very narrow spectral features, we also analyze the ringdown signal using the filter diagonalization method \cite{Mandelshtam1997:harminv} via the Harminv software \cite{website:harminv}, which extracts the decay rates and frequencies of the high-Q cavity modes.

{
\newcommand{\scaler}{0.5}
\begin{figure}
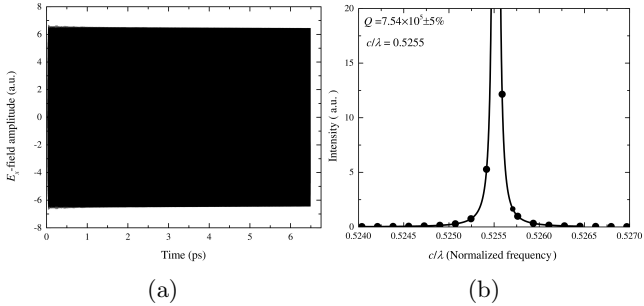

     \centering

    \begin{subfigure}[b]{\scaler\linewidth}
      \centering
      \includegraphics[width=0.99\textwidth]{figure06a}
      \caption{}
      \label{fig:probeSignal-TimeDomain}
    \end{subfigure}%
    \begin{subfigure}[b]{\scaler\linewidth}
      \centering
      \includegraphics[width=0.99\textwidth]{figure06b}
      \caption{}
      \label{fig:probeSignal-FFT}
    \end{subfigure}%

    \caption{
      Amplitude of the electric field $E_{x}$ as a function of time (a) and the corresponding fast Fourier transform (b), near the defect in a woodpile with $n_{def}=n_{wp}=3.3$ and $n_{bf}=1$, after an initial excitation pulse in the X direction.
      Using Harminv \cite{Mandelshtam1997:harminv,website:harminv}, we find a resonance peak at $ \lambda_{0}^{res} \simeq \ExampleLambdaNanometers nm $ with a Q-factor of $ \ExampleQfactor \pm 5\% $.
      It corresponds to the fourth column of \tref{tab:tabular-DefectSizes}.
    }
    \label{fig:probeSignal}
\end{figure}
}

\subsection{Calculating the effective mode volume $V_{eff}$}

Having determined the resonant frequency, we can visualise a cavity mode on resonance using single frequency "snapshots". We illustrate the confinement of the electric field energy density distribution in each plane y-z, x-z, and y-x in \fref{fig:figure-energySnapshots-D1} for the defect D1. It shows x, y, and z projections of the woodpile superimposed over the electric field energy density distributions ($\varepsilon\cdot|E|^{2} = \varepsilon\cdot(|E_x|^{2}+|E_y|^{2}+|E_z|^{2})$, although the plotted values are actually $\varepsilon_{r}\cdot|E|^{2}$ with $\varepsilon_{r}=\varepsilon/\varepsilon_{0}$) of a defect for an $E_{x}$ oriented dipole mode.

In all cases the field is strongly localised to the defect, with some spread to the high refractive index links nearby. 
The FDTD algorithm allows a computation of the effective mode volume of the cavity modes with high Q-factor, which is the 3D PhC cavity mode that survives after a sufficiently long period of time.
We do this by creating a sequence of resonant frequency "snapshots" through the resonant mode, covering all grid points, then digitally summing the fields using the definition of the effective mode volume $V_{eff}$:
\begin{equation}
V_{eff}=\frac{\iiint{ \varepsilon(r)\left|E(r)\right|^{2}d^{3}r }}{\left[\varepsilon(r)\left|E(r)\right|^{2}\right]_{max}}
\label{eq:ModeVolume}
\end{equation}
We see that $V_{eff}$ is given by the spatial integral of the electric field intensity for the cavity mode, divided by its maximum value.
The $\varepsilon(r)$ in \eref{eq:ModeVolume} is equal to $\frac{n^{2}(r)}{\mu_{r}}\varepsilon_{0}$ and therefore proportional to the square of the spatially dependent refractive index $n(r)$.
A figure of merit is the dimensionless effective volume $f_{opt}$, which is the effective cavity mode volume $V_{eff}$ normalised to the cubic wavelength of the resonant mode $(\lambda/n)^{3}$ in a medium of refractive index $n$, defined as:
\begin{equation}
f_{opt} = \frac{V_{eff}}{(\lambda/n)^{3}}
\label{eq:normalizedModeVolume}
\end{equation}

Since $E_{max}$ is proportional to the inverse square root of the mode volume ($V_{eff}^{-1/2}$), the field coupling strength can be enhanced by reducing $V_{eff}$. 

{
\newcommand{\widthfraction}{0.25}

\begin{figure*}
  \centering

  \begin{subfigure}[b]{\widthfraction\linewidth{}}
    \centering
    \includegraphics[width=.99\linewidth]{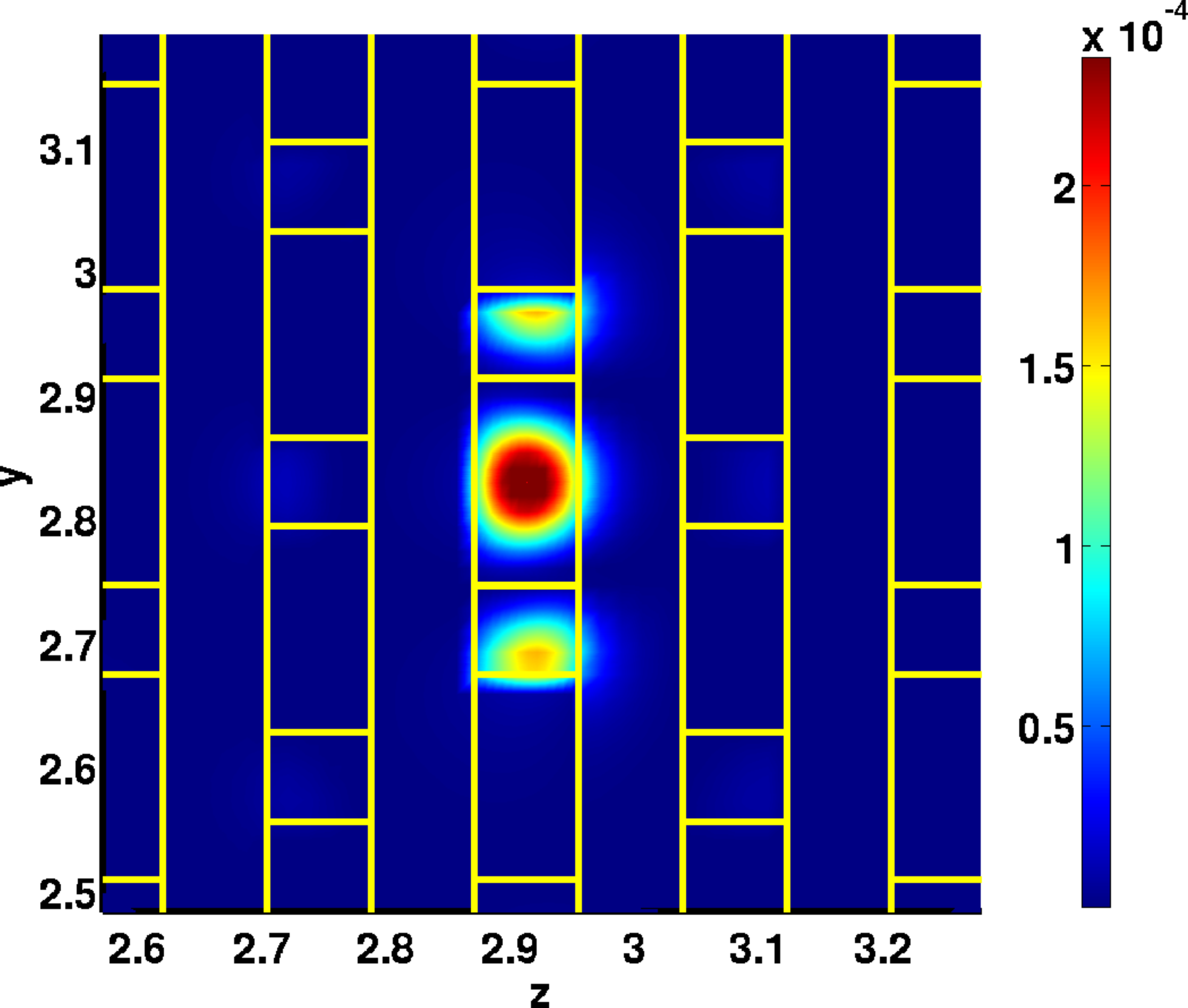}
    \caption{}
    \label{fig:D1-Ex-2D}
  \end{subfigure}%
  \begin{subfigure}[b]{\widthfraction\linewidth{}}
    \centering
    \includegraphics[width=.99\linewidth]{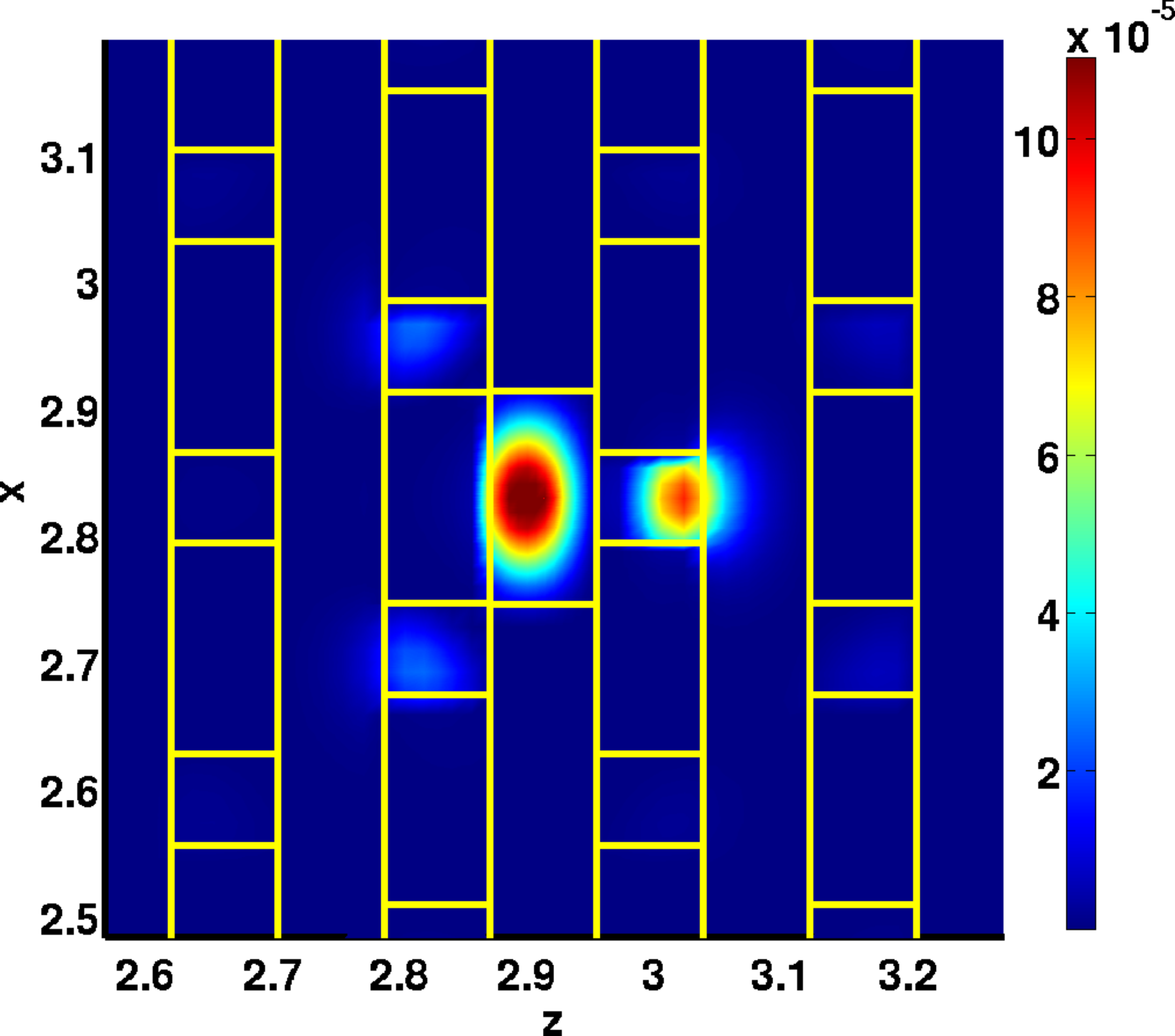}
    \caption{}
    \label{fig:D1-Ey-2D}
  \end{subfigure}%
  \begin{subfigure}[b]{\widthfraction\linewidth{}}
    \centering
    \includegraphics[width=.99\linewidth]{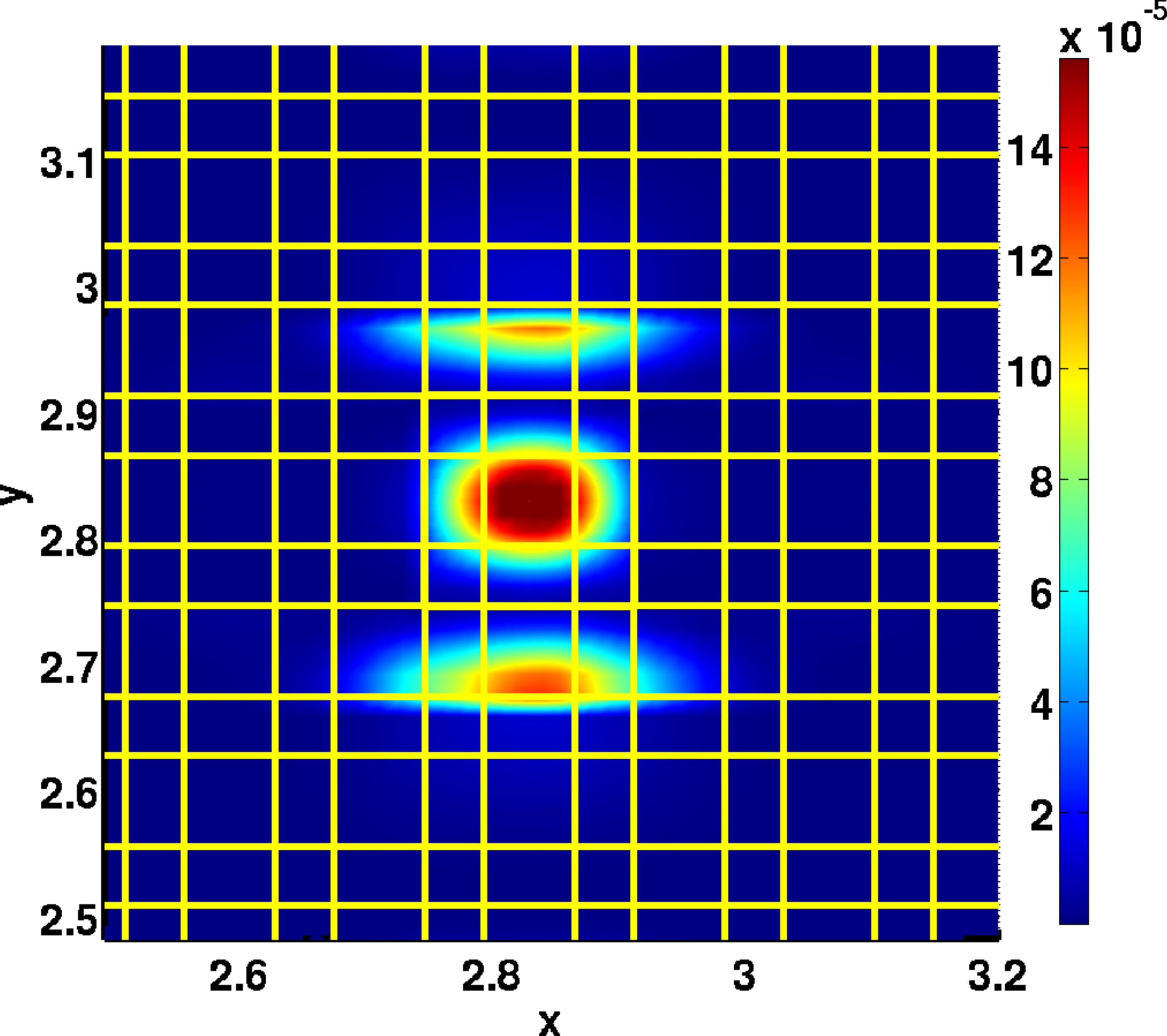}
    \caption{}
    \label{fig:D1-Ez-2D}
  \end{subfigure}

  \begin{subfigure}[b]{\widthfraction\linewidth{}}
    \centering
    \includegraphics[width=.99\linewidth]{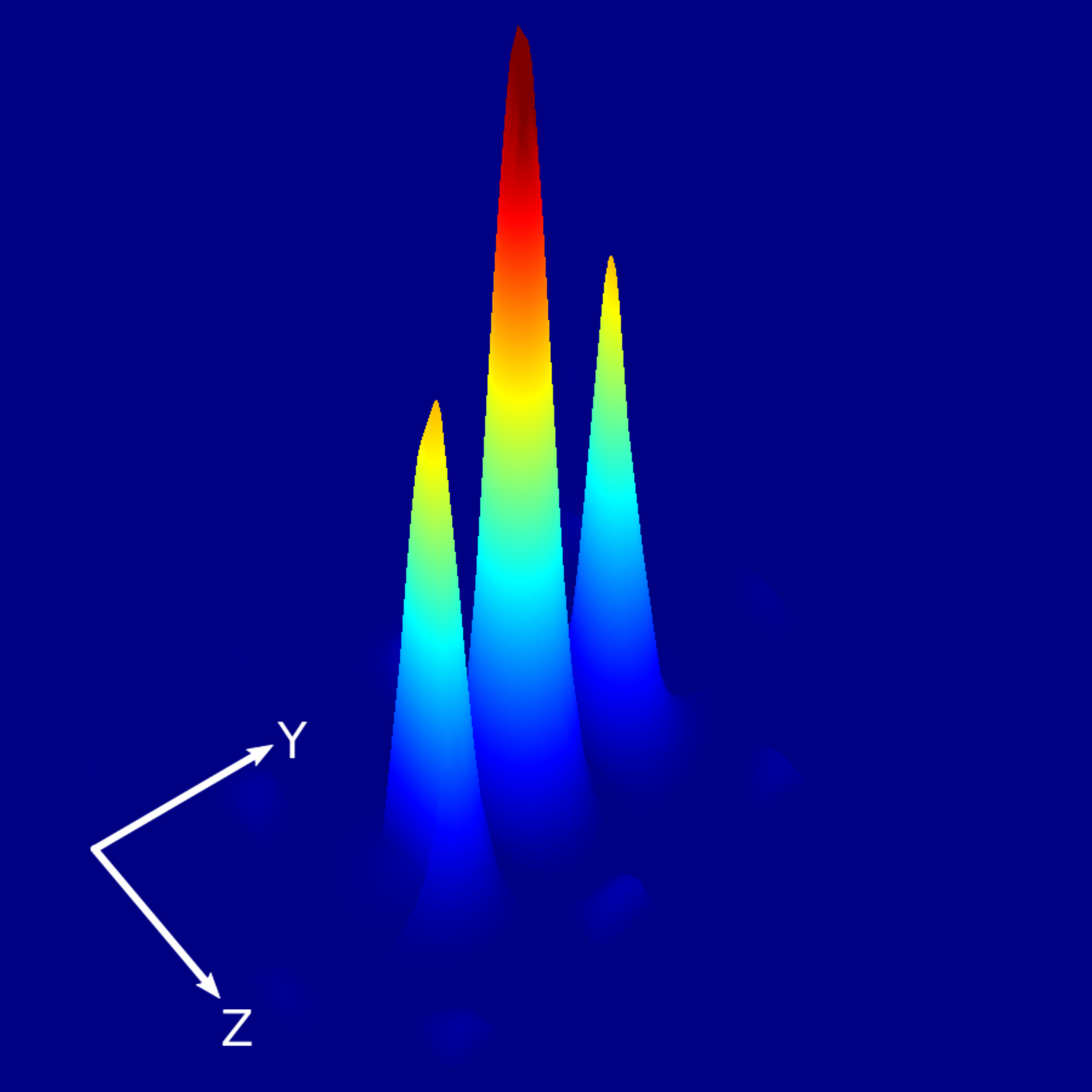}
    \caption{}
    \label{fig:D1-Ex-3D-v1}
  \end{subfigure}%
  \begin{subfigure}[b]{\widthfraction\linewidth{}}
    \centering
    \includegraphics[width=.99\linewidth]{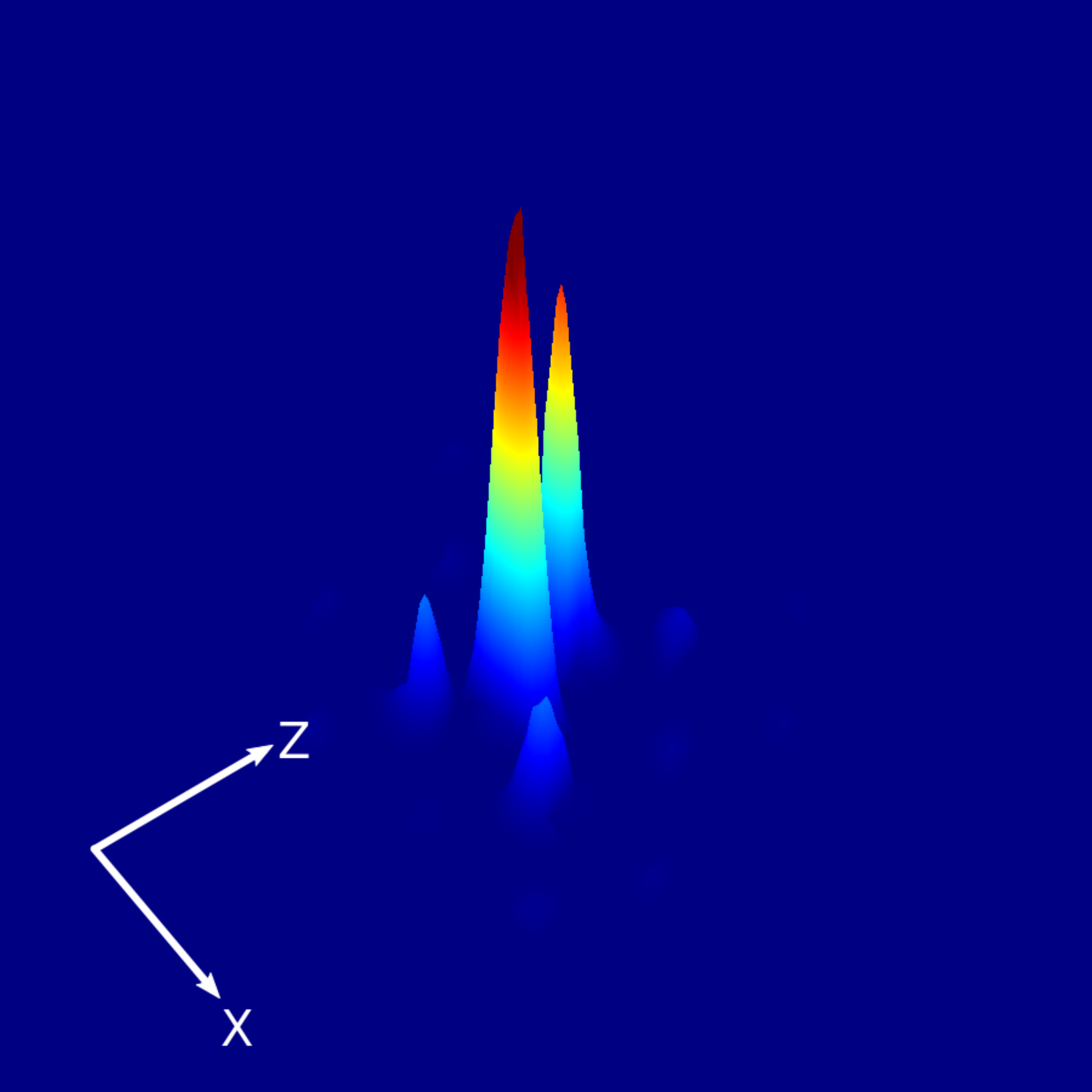}
    \caption{}
    \label{fig:D1-Ey-3D}
  \end{subfigure}%
  \begin{subfigure}[b]{\widthfraction\linewidth{}}
    \centering
    \includegraphics[width=.99\linewidth]{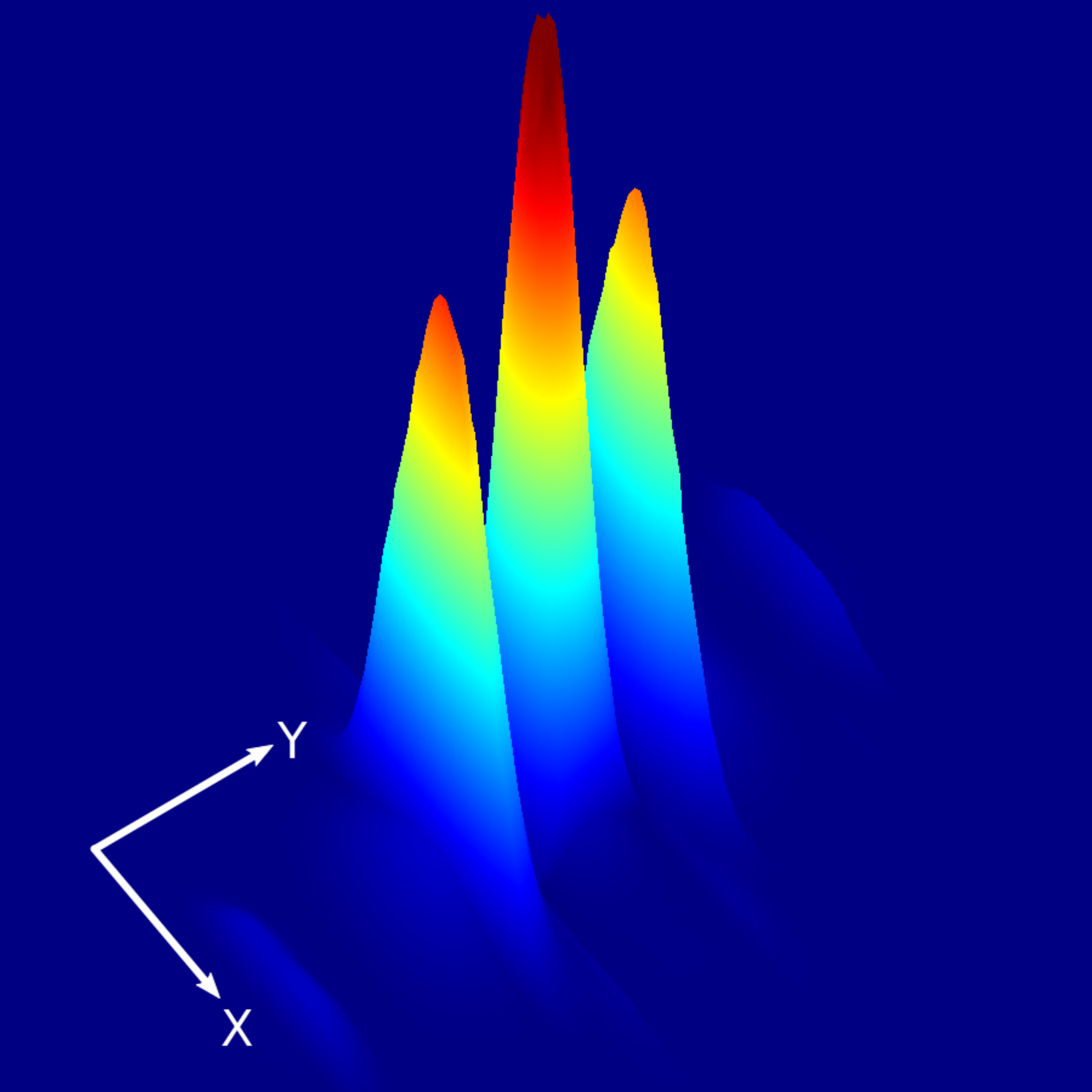}
    \caption{}
    \label{fig:D1-Ez-3D}
  \end{subfigure}

  \caption{
  The resonant frequency snapshots of a defect cavity mode energy density distribution (\ElectromagneticEnergy) in the central X (a,d), Y (b,e), and Z (c,f) planes for a woodpile with $ n_{def} = n_{wp} = 3.3 $, $ n_{bf} = 1 $ and a D1 defect, after an initial broadband gaussian modulated sinewave excitation pulse.
  High Q-factor resonant modes at
  $\VerticalPeriod/\lambda_{0} \simeq 0.5255 $ with an $E_{x}$-oriented dipole source (a,d),
  $\VerticalPeriod/\lambda_{0} \simeq 0.5058 $ with an $E_{y}$-oriented dipole source (b,e) and
  $\VerticalPeriod/\lambda_{0} \simeq 0.5300 $ with an $E_{z}$-oriented dipole source (c,f).
  }
  \label{fig:figure-energySnapshots-D1}
\end{figure*}
}


\subsection{Estimating the Purcell enhancement $F_{p}$ and the coupling strength $g_{R}$}


In order to evaluate the usefulness of the simulated cavities for quantum information applications, we estimate how strongly a quantum emitter, which can be considered as a transition dipole, placed inside it will interact with the vacuum field created by the cavities.
A strong interaction means it will be possible to entangle photon states with quantum emitter states.

In this paper, we consider diamond NV-centres as an example quantum emitter.
Diamond NV-centres are interesting due to their stability as a single photon source, their long spin decoherence time ($\sim ms$) and their ability to generate indistinguishable photons at low-temperatures \cite{Sipahigil2012, Bernien2012}.
Additionally, their availability in the form of defects in nanocrystals makes it possible to embed them within photonic crystals fabricated using a different material, for instance 3D photonic crystals made via direct laser writing.

\subsubsection{Emission spectrum and coupling rate of a coupled dipole-cavity system: \label{section:spectrum-theory}}

When the cavity mode radial frequency $\omega_{cav}$ is equal
to the transition dipole radial frequency $\omega_{os}$ (corresponding
to the energy difference between the ground and excited state of the
quantum emitter) and if the dipole has the same polarization
as the cavity mode, the luminescence spectrum for such a system
is\cite{Carmichael1989,Andreani1999}:

\begin{equation}
S(\omega)\propto\left|\frac{\Omega_{+}-\omega_{0}+i\frac{\kappa}{2}}{\omega-\Omega_{+}}-\frac{\Omega_{-}-\omega_{0}+i\frac{\kappa}{2}}{\omega-\Omega_{-}}\right|\label{eq:spectrum}
\end{equation}
where $\omega_{0}=\omega_{cav}=\omega_{os}$ and:

\begin{equation}
\Omega_{\pm}=\omega_{0}-\frac{i}{4}(\kappa+\gamma)\pm\sqrt{g_{R}^{2}-\left(\frac{\kappa-\gamma}{4}\right)^{2}}\label{eq:eigenfrequencies-on-resonance}
\end{equation}

$\kappa=\Delta\omega=\frac{\omega_{cav}}{Q}$ is the \emph{full width at half-maximum} (FWHM) of the cavity mode, corresponding to the rate of escape of photons from the cavity, and $\gamma$ is the FWHM of the emission spectrum of the \emph{quantum emitter dipole} without any cavity (equivalent to its \emph{spontaneous emission rate} assuming no other broadening processes).
For diamond NV-centres, we take the emission rate into the zero-phonon line which is resonant with the cavity: $\gamma=\gamma^{ZPL}=2\pi\cdot3.3\cdot10^{-3}GHz$ \cite{Ho2011:IEEE_JQE, McCutcheon2008, Wolters2010a, Santori2010a, Collins1983, Beveratos2002}.

The \emph{dipole-cavity coupling rate} $g_{R}$ is given by\cite{Andreani1999}:

\begin{equation}
g_{R}=\sqrt{\frac{\pi\cdot e_{0}^{2}\cdot f^{EG}}{4\cdot\pi\cdot\varepsilon_{0}\cdot n_{def}^{2}\cdot V_{eff}\cdot m_{e}}}\label{eq:coupling-rate}
\end{equation}
$e_{0}$ and $m_{e}$ are the elementary charge and mass of an electron respectively, $\varepsilon_{0}$ is the vacuum permittivity, $n_{def}$ is the refractive index of the defect, where the dipole is located and $V_{eff}$ is the mode volume of the cavity mode.
$f^{EG}$ is the oscillator strength of the dipole given by:

\begin{equation}
f^{EG}=\frac{2\cdot m_{e}\cdot\omega_{os}\cdot d_{EG}^{2}}{e_{0}^{2}\cdot\hbar}\label{eq:oscillator-strength}
\end{equation}
where $d_{EG}$ is the dipole moment of the quantum emitter:

\begin{equation}
d_{EG}=\sqrt{\frac{3\cdot\pi\cdot\varepsilon_{0}\cdot c_{0}^{3}\cdot\hbar}{n_{os}\cdot\omega_{os}^{3}}\times\gamma}\label{eq:dipole-moment}
\end{equation}
$n_{os}$ is the refractive index of the material directly surrounding
the quantum emitter. In the case of NV-centres, this is diamond with $n_{os}=2.4$.

Equations \eref{eq:coupling-rate}, \eref{eq:oscillator-strength} and
\eref{eq:dipole-moment} lead to:

\begin{equation}
g_{R}=\left(\frac{3Q}{4\pi^{2}V_{eff}}\times\frac{\lambda_{os}^{3}}{n_{def}^{2}n_{os}}\times\frac{\kappa\gamma}{4}\right)^{\frac{1}{2}}\label{eq:coupling-rate-2}
\end{equation}

\subsubsection{Strong/weak coupling limit:}

As illustrated in \cite{Khitrova2006}, the luminescence spectrum
$S(\omega)$ given in \eref{eq:spectrum} corresponds to a
superposition of two peaks, with central frequencies $\Re(\Omega_{+})$
and $\Re(\Omega_{-})$ and whose linewidths are:

\begin{eqnarray}
\gamma_{+} & \simeq & -2\Im(\Omega_{+}) \\
           & =      & -2\Im\left(\omega_{0}-\frac{i}{4}(\kappa+\gamma)+\sqrt{g_{R}^{2}-\left(\frac{\kappa-\gamma}{4}\right)^{2}}\right)\label{eq:FWHM-p}\\
\gamma_{-} & \simeq & -2\Im(\Omega_{-}) \\
           & =      & -2\Im\left(\omega_{0}-\frac{i}{4}(\kappa+\gamma)-\sqrt{g_{R}^{2}-\left(\frac{\kappa-\gamma}{4}\right)^{2}}\right)\label{eq:FWHM-m}
\end{eqnarray}

When $4g_{R}<\left|\kappa-\gamma\right|$, the general two peak spectrum collapses to
a single peak at frequency $\omega_{0}$. This is called the weak
coupling regime. When $4g_{R}>\left|\kappa-\gamma\right|$, there are two separate
frequency components at $\omega_{\pm}=\omega_{0}\pm\sqrt{g_{R}^{2}-\left(\frac{\kappa-\gamma}{4}\right)^{2}}$ and this is called the strong coupling regime.
However, this only leads to two separate peaks in the spectrum for:

\begin{equation}
\frac{4g_{R}}{\kappa+\gamma}>1
\end{equation}
This is the commonly used condition for strong coupling.
The weak and strong coupling regimes are characterised by the reversibility of the emission. In the weak coupling regime, photons emitted by the quantum emitter
are very unlikely to be reabsorbed by it, i.e. the emission is irreversible.
The converse is true for the strong coupling regime where emitted photons are very likely to be reabsorbed by the quantum emitter, i.e. the emission is reversible.

\subsubsection{Spontaneous emission modification: Purcell factor:}

In the weak coupling regime, for varying values of $\kappa$, $\gamma_{+}$
tends to stay close to $\gamma$, while $\gamma_{-}$ stays close
to $\kappa$. $\gamma_{+}$ therefore corresponds to the emission rate of the coupled
dipole, while $\gamma_{-}$ corresponds to the decay rate of the
coupled cavity mode. The modification of the spontaneous emission rate $\gamma$ can thus
be quantified by the so-called Purcell factor:

\begin{equation}
F_{p}=\frac{\gamma_{+}}{\gamma}\label{eq:Purcell-factor}
\end{equation}

\subsubsection{Purcell factor approximation in the weak coupling regime:}

In the weak coupling regime, when $4g_{R}\ll\left|\kappa-\gamma\right|$,
but also when $\kappa\gg\gamma$ (which is usually the case) and $4g^{2}\gg\kappa\gamma$,
$\Omega_{+}$ and $\Omega_{-}$ can be simplified to\cite{Andreani1999}:

\begin{eqnarray}
\Omega_{+} & \simeq & \omega_{0}-i\left(\frac{\gamma}{2}+\frac{2g_{R}^{2}}{\kappa}\right)=\omega_{0}-i\left(\frac{2g_{R}^{2}}{\kappa}\right)\label{eq:Omega-p-simple}\\
\Omega_{-} & \simeq & \omega_{0}-i\left(\frac{\kappa}{2}-\frac{2g_{R}^{2}}{\kappa}\right)\label{eq:Omega-m-simple}
\end{eqnarray}

Equations \eref{eq:Purcell-factor}, \eref{eq:FWHM-p}, \eref{eq:Omega-p-simple}
and \eref{eq:coupling-rate-2} then lead to:

\begin{equation}
F_{p}\simeq\frac{4g_{R}^{2}}{\kappa\gamma}=\frac{3Q}{4\pi^{2}V_{eff}}\times\frac{\lambda_{os}^{3}}{n_{def}^{2}n_{os}}
\label{eq:Fp-long}
\end{equation}

When the nano-diamond is small ($\ll \lambda$) and embedded in a defect, $n_{os} \sim n_{def}$.
This allows us to further simplify \eref{eq:Fp-long} to the well known approximation of the Purcell factor in the weak coupling regime:

\begin{equation}
F_{p}\simeq\frac{3Q\left(\frac{\lambda_{os}}{n_{def}}\right)^{3}}{4\pi^{2}V_{eff}}
\label{eq:Fp-short}
\end{equation}

This is the approximation we used for the values presented in tables \tref{tab:tabular-DefectSizes} and \tref{tab:tabular-AirBuffers} and \fref{fig:PurcellFactor-plot}.
However, due to the strong coupling we obtained in most cases, it is not a valid indicator of the spontaneous emission modification.
In the strong coupling region the peak splitting leads to an oscillating probability of emission which decays at a rate largely determined by the cavity lifetime.


\section{Results}

\subsection{Results for different defect sizes}

First, we look at the defects without air buffers D0, D1 and D2. \Tref{tab:tabular-DefectSizes} and \fref{fig:table-plots} show the corresponding results.
Simulations for the defect D0 in the $E_{x}$ direction showed no clearly distinguishable resonance modes, which is why its entries in \tref{tab:tabular-DefectSizes} are marked \textit{n/a}.
The defect D1 gives the best results for an excitation in the $E_{x}$ direction with a Q-factor $ Q \simeq \SbXQ $ and a mode volume $ V_{eff} \simeq \SbXMVnormalized $.

In terms of coupling strength, all defects without air buffer exhibit strong coupling capabilities, except D0 in the $E_{x}$ and $E_{y}$ directions.
Even in the $E_{z}$ direction, it barely exceeds the strong coupling condition with $ g_{R}/(2\pi) \simeq 5.85 GHz $ and $ 4 g_{R} / (\kappa+\gamma) = 1.66 $.
It does however offer a spontaneous emission enhancement of $ F_{p} = 28.5 $.

D1 shows the best coupling strength with $ g_{R}/(2\pi) \simeq 6.37 GHz $ and $ 4 g_{R} / (\kappa+\gamma) = 40.57 $.
For both D1 and D2, when moving from $E_{x}$ excitations to $E_{y}$ and $E_{z}$ excitations, the Q-factors become smaller and the mode volumes larger, i.e. the confinement properties get worse.
We think this is because dipoles in the $E_y$ and $E_z$ directions emit most of their energy in the XZ and XY plane respectively, but since the emitter is placed between two rods in the X direction, there is nothing preventing energy escaping along the X direction, apart from the defect box itself. 

\subsection{Results for different air buffer sizes}

\begin{figure}
  \centering

  \newcommand{\TablePlotSize}{width=1\linewidth,height=0.25\textheight}
  \newcommand{\TablePlotSubFigureWidth}{1\linewidth}
  
  \begin{subfigure}[b]{\TablePlotSubFigureWidth}
    \centering
    \includegraphics[height=0.2\textheight]{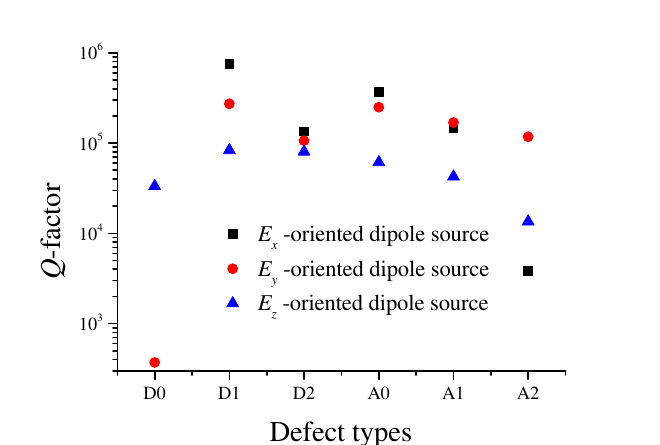}
    \caption{}
    \label{fig:Qfactor-plot}
  \end{subfigure}%
  
  \begin{subfigure}[b]{\TablePlotSubFigureWidth}
    \centering
    \includegraphics[height=0.2\textheight]{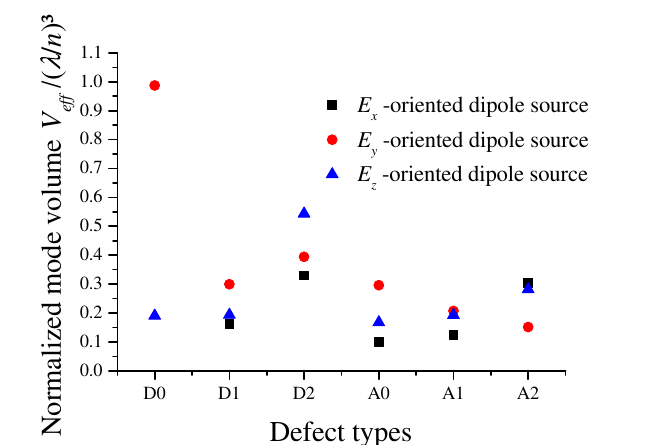}
    \caption{}
    \label{fig:ModeVolume-plot}
  \end{subfigure}%

  \begin{subfigure}[b]{\TablePlotSubFigureWidth}
    \centering
    \includegraphics[height=0.2\textheight]{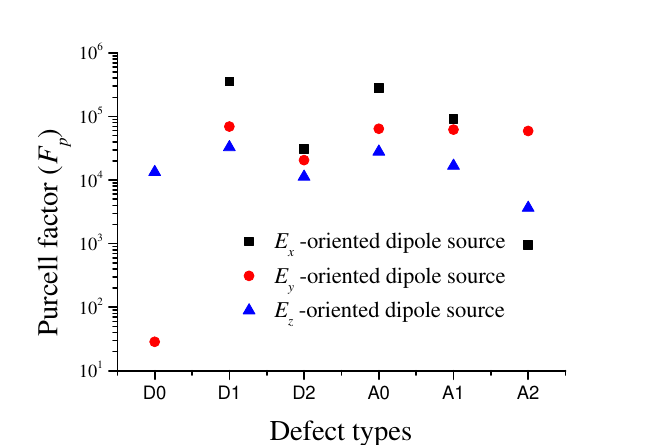}
    \caption{}
    \label{fig:PurcellFactor-plot}
  \end{subfigure}%
  
  \begin{subfigure}[b]{\TablePlotSubFigureWidth}
    \centering
    \includegraphics[height=0.2\textheight]{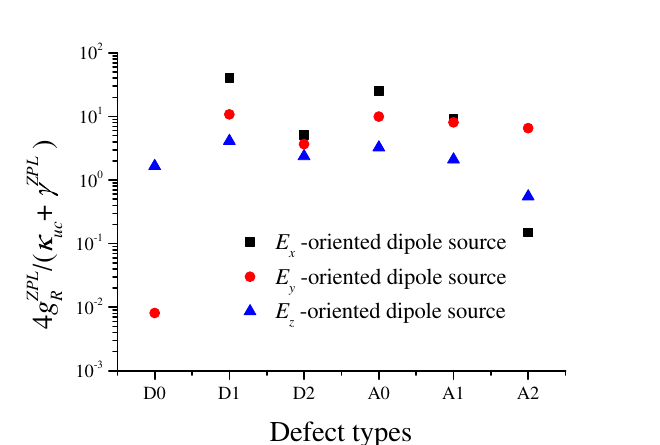}
    \caption{}
    \label{fig:CouplingStrength-plot-log}
  \end{subfigure}%

  \caption{
    Q-factors (a), mode volumes (b) and Purcell factors (c) of the resonant modes for the six defects studied in this paper.
    Additionally (d) shows the strong coupling condition $\frac{4g_{R}}{\kappa_{uc}+\gamma}$. The strong coupling regime is attained when it is larger than 1.
  }
  \label{fig:table-plots}
\end{figure}

We now concentrate on the defect D1, which gave the best results.
As can be seen on the corresponding energy snapshots in \fref{fig:figure-energySnapshots-D1}, the energy tends to leak into the rods touching the defect.
In order to improve the confinement, i.e. reduce the leakage, we therefore now add the previously mentioned cuboid air buffers around the defect to disconnect it from the surrounding rods.
\Tref{tab:tabular-AirBuffers} and figures \ref{fig:figure-energySnapshots-A0-A1-A2} and \ref{fig:table-plots} show the results for the defects with air buffers A0, A1 and A2.
This essentially corresponds to "cutting" into the rods by $ 1/4 \cdot \RodWidth $, $ 1/2 \cdot \RodWidth $ and $ 3/4 \cdot \RodWidth $. Since $b_z = d_z = 0.25 \cdot \VerticalPeriod = \RodHeight $ and there is a rod touching the top of the defect as can be seen in figures \ref{fig:A0}, \ref{fig:A1}, \ref{fig:A2}, the structure remains contiguous and could be fabricated directly or as an inverse for backfilling with high refractive index material.

Figures \ref{fig:figure-energySnapshots-D1-A0-A1-A2} and \ref{fig:central-cut} show that the two peaks on the side of the main energy peak in the YZ plane are indeed reduced by the air buffers, the most efficient being the air buffer used in A2. However, while the mode volume also dropped as expected for the $E_{x}$ excitation in the A0 and A1 defects, it instead increased for the A2 defect.
A2 is also the only one of the defects with air buffers not suitable for strong coupling, with $ 4 g_{R} / (\kappa+\gamma) = 0.15 $ for an $E_{x}$ excitation and $ 4 g_{R} / (\kappa+\gamma) = 0.85 $ for an $E_{z}$ excitation.
The corresponding emission enhancements are $ F_{p} = 9.45 \times{} 10^{2} $ and $ F_{p} = 8.58 \times{} 10^{3} $.

{
\newcommand{\widthfraction}{0.24}
\newcommand{\subsubsize}{0.9}

\begin{figure*}
  \centering

  \begin{subfigure}[b]{\widthfraction\linewidth{}}
    \centering
    \includegraphics[width=.99\linewidth]{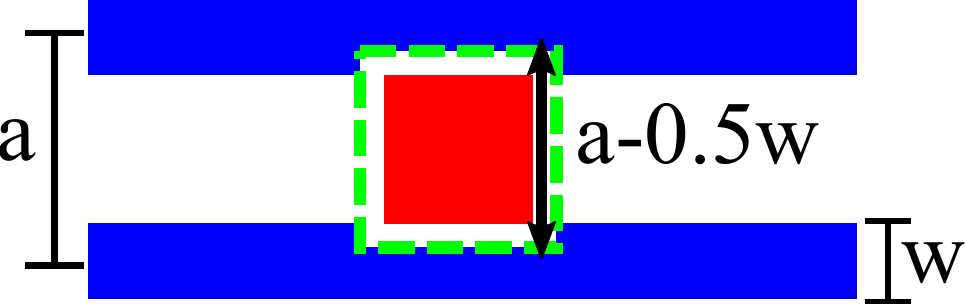}
    \caption{A0}
    \label{fig:A0-2D-illustration}
  \end{subfigure}%
  \begin{subfigure}[b]{\widthfraction\linewidth{}}
    \centering
    \includegraphics[width=.99\linewidth]{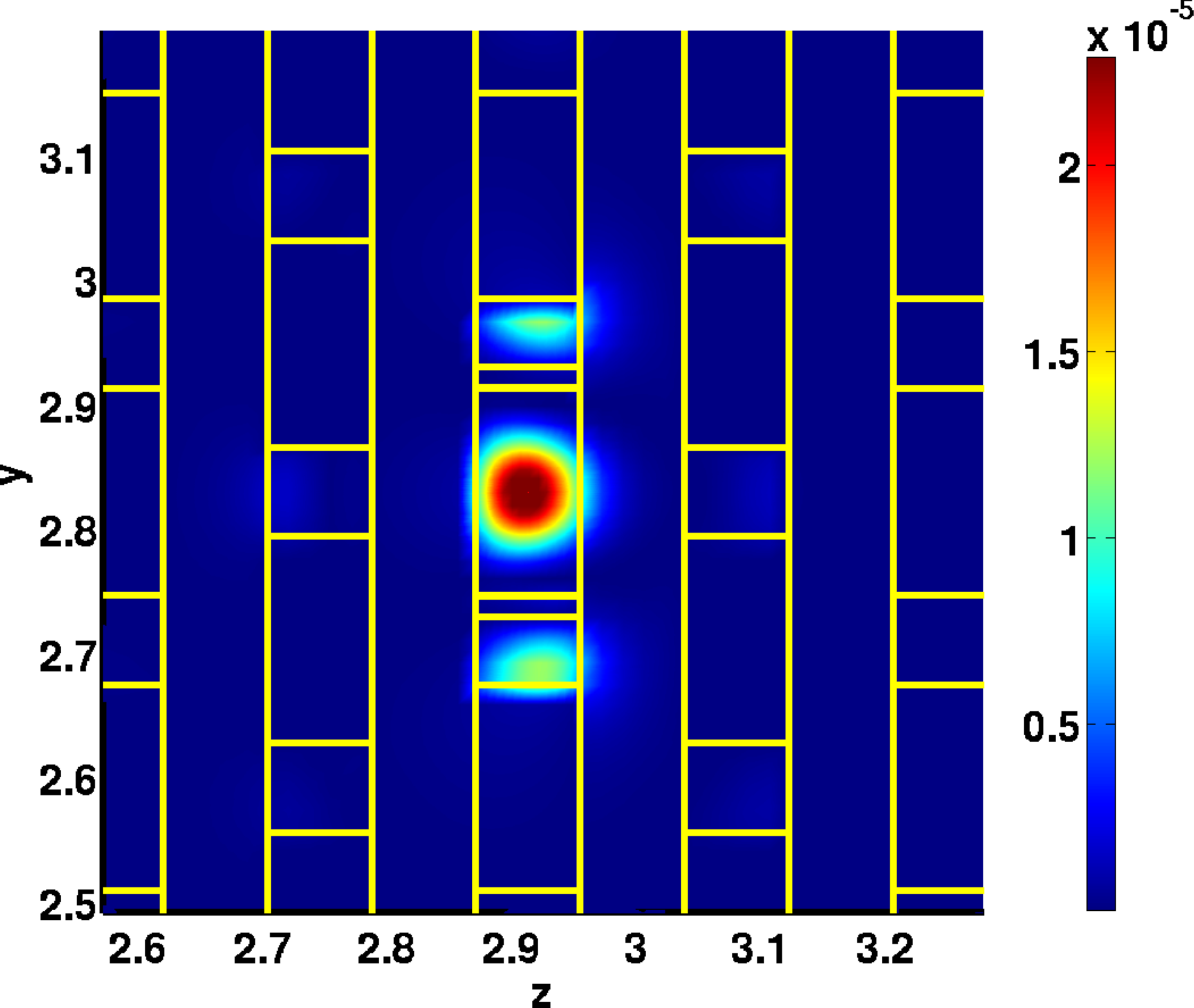}
    \caption{A0}
    \label{fig:A0-Ex-2D}
  \end{subfigure}%
  \begin{subfigure}[b]{\widthfraction\linewidth{}}
    \centering
    \includegraphics[width=.99\linewidth]{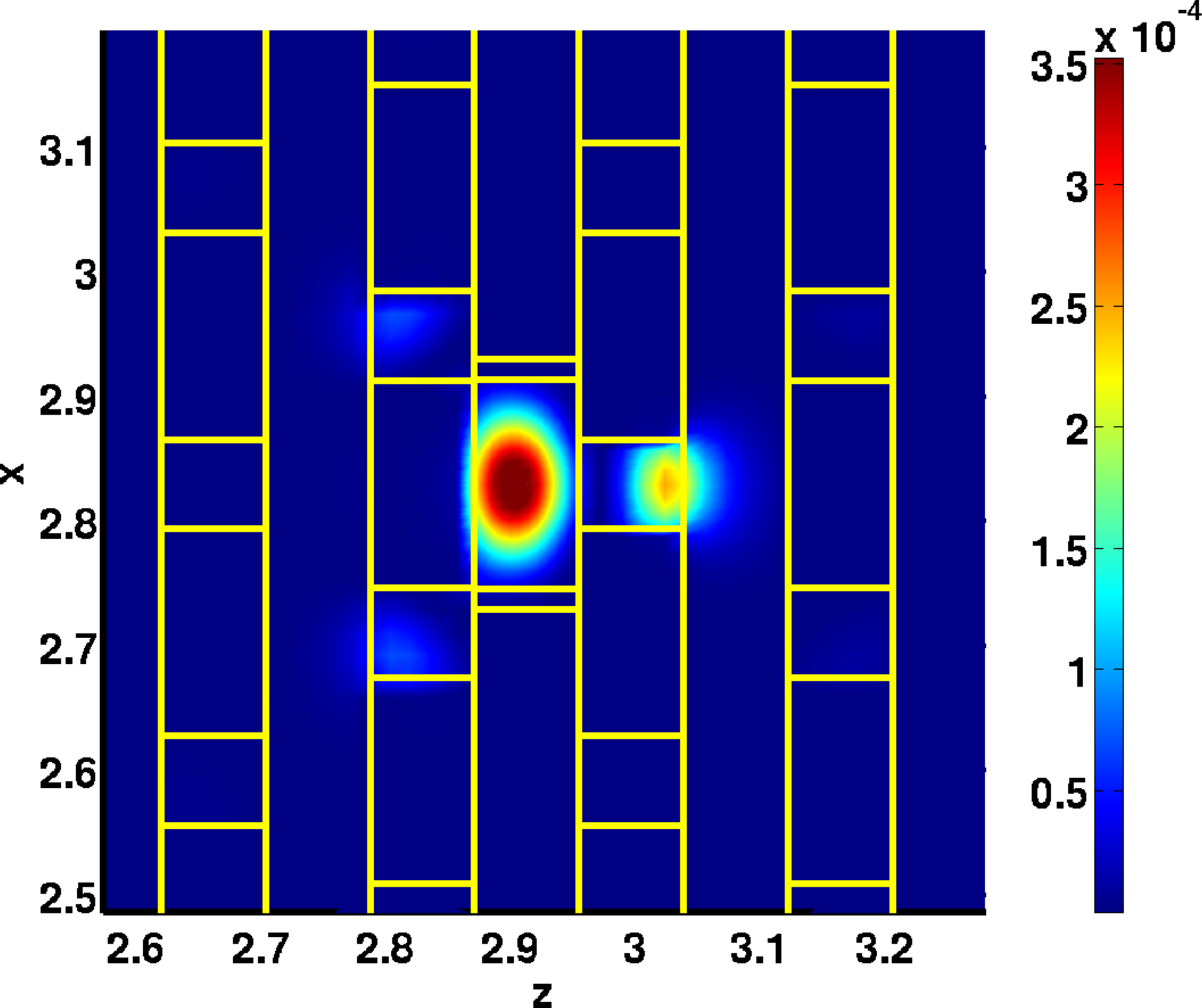}
    \caption{A0}
    \label{fig:A0-Ey-2D}
  \end{subfigure}
  \begin{subfigure}[b]{\widthfraction\linewidth{}}
    \centering
    \includegraphics[width=.99\linewidth]{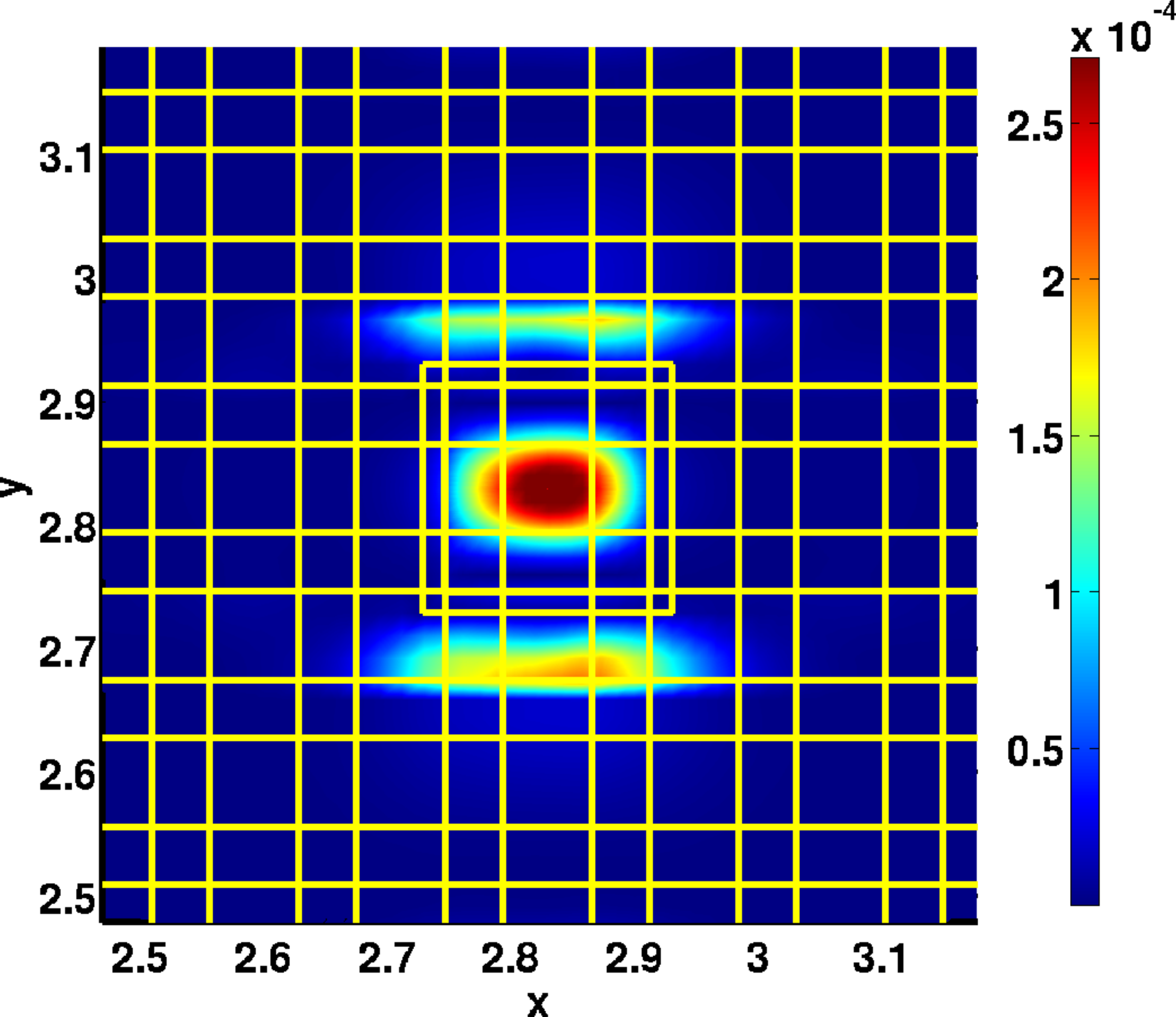}
    \caption{A0}
    \label{fig:A0-Ez-2D}
  \end{subfigure}

  \begin{subfigure}[b]{\widthfraction\linewidth{}}
    \centering
    \includegraphics[width=.99\linewidth]{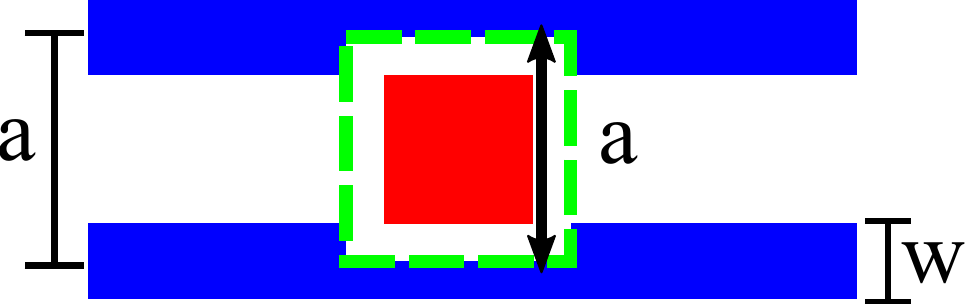}
    \caption{A1}
    \label{fig:A1-2D-illustration}
  \end{subfigure}%
  \begin{subfigure}[b]{\widthfraction\linewidth{}}
    \centering
    \includegraphics[width=.99\linewidth]{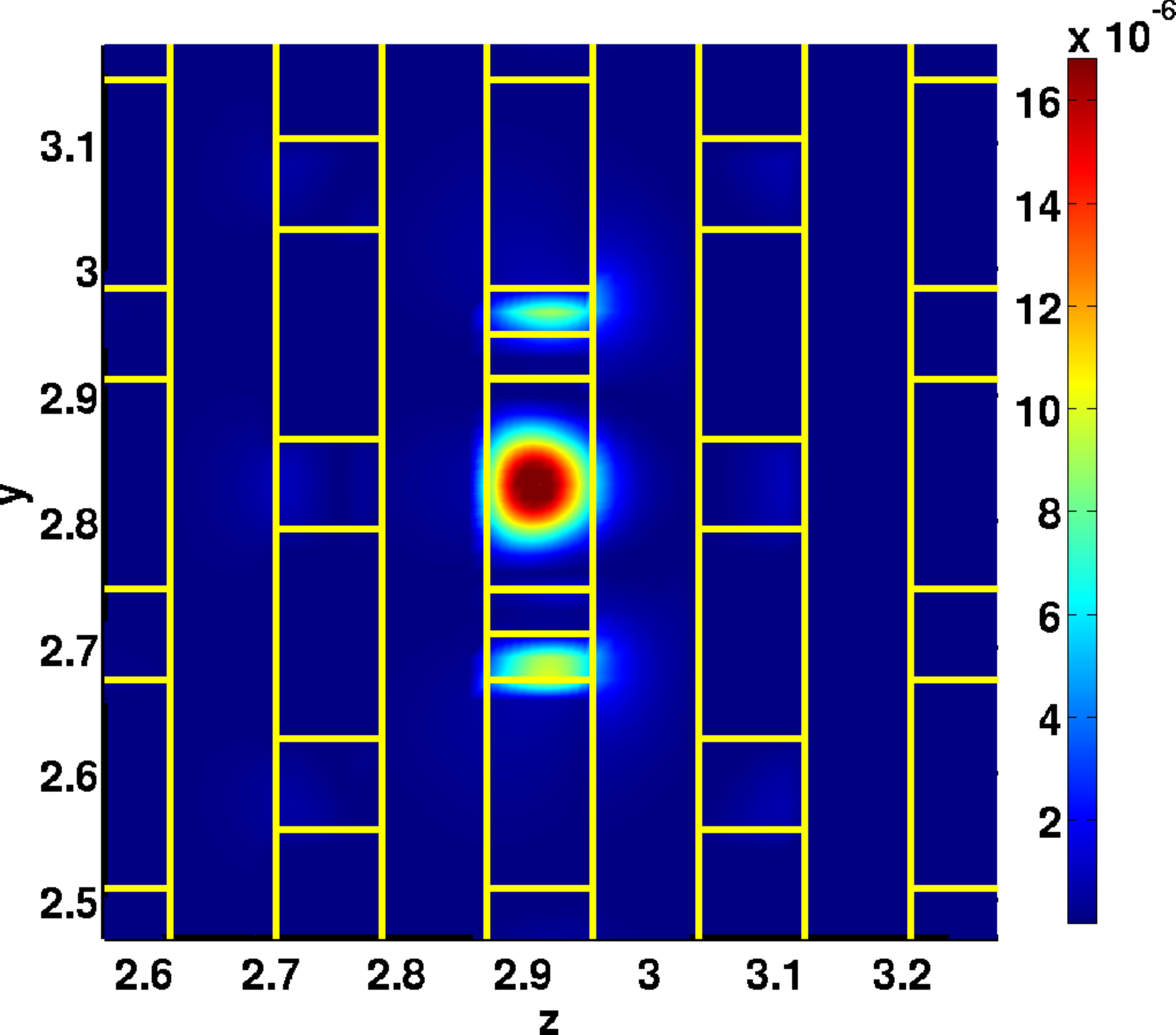}
    \caption{A1}
    \label{fig:A1-Ex-2D}
  \end{subfigure}%
  \begin{subfigure}[b]{\widthfraction\linewidth{}}
    \centering
    \includegraphics[width=.99\linewidth]{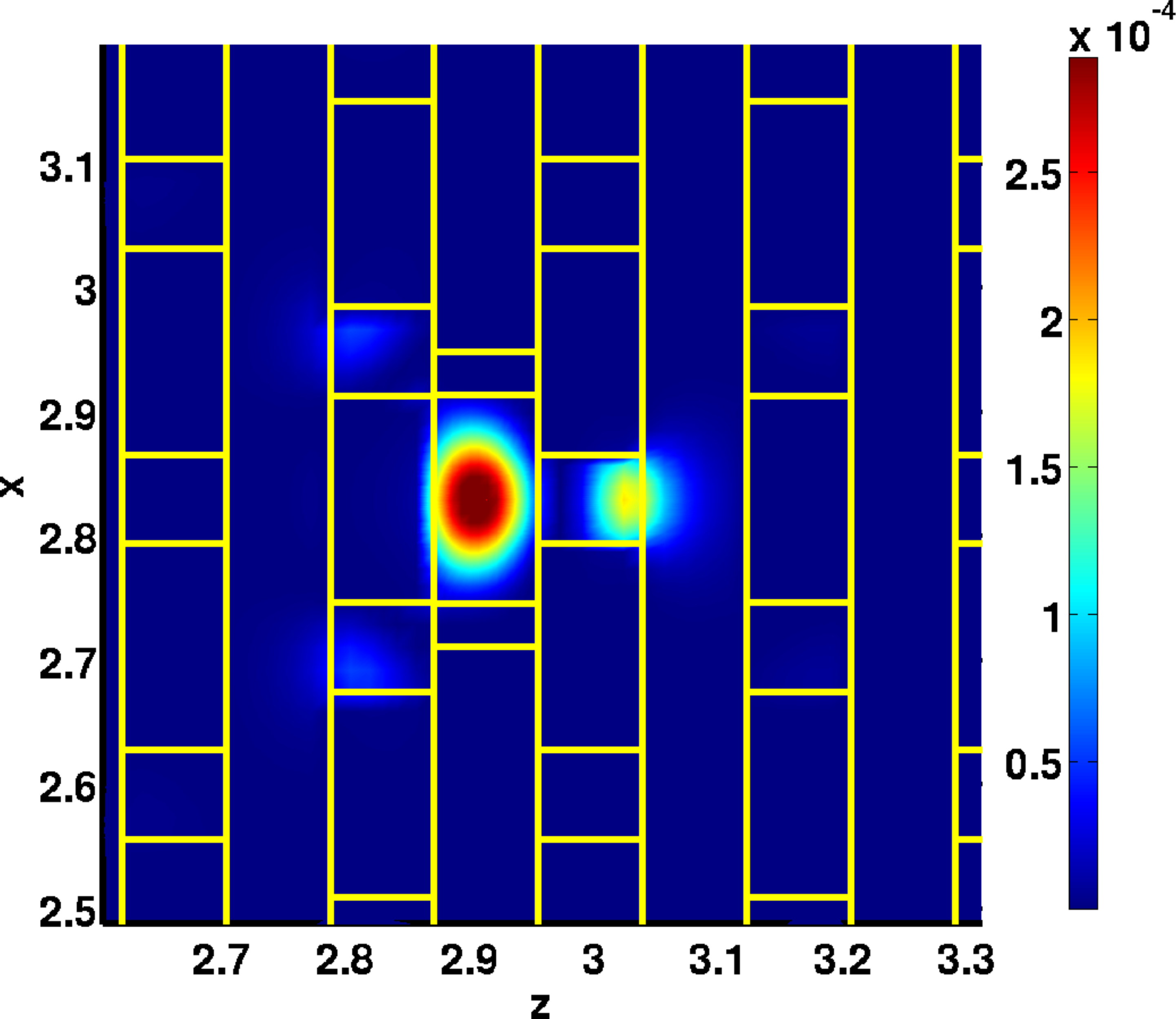}
    \caption{A1}
    \label{fig:A1-Ey-2D}
  \end{subfigure}
  \begin{subfigure}[b]{\widthfraction\linewidth{}}
    \centering
    \includegraphics[width=.99\linewidth]{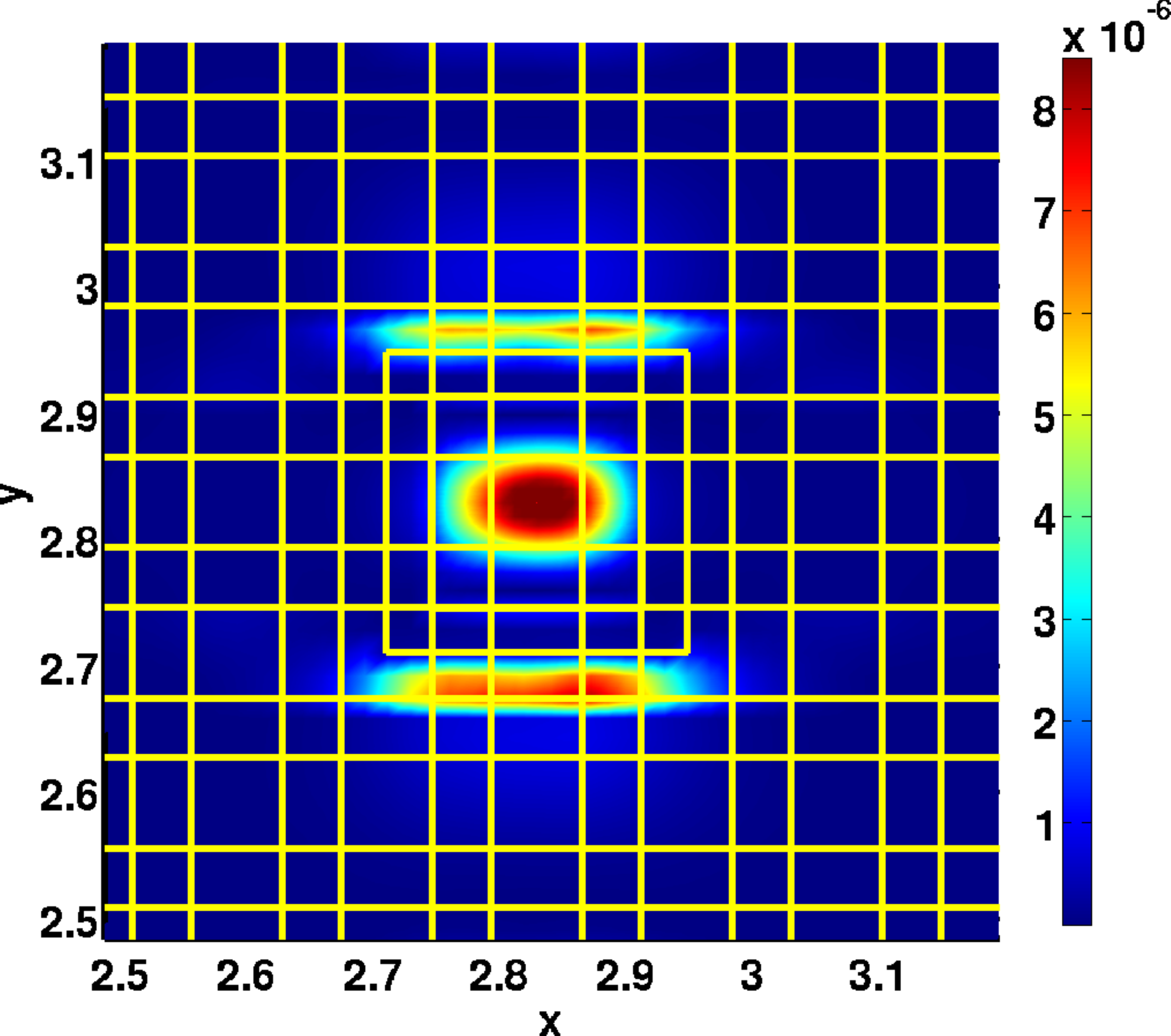}
    \caption{A1}
    \label{fig:A1-Ez-2D}
  \end{subfigure}

  \begin{subfigure}[b]{\widthfraction\linewidth{}}
    \centering
    \includegraphics[width=.99\linewidth]{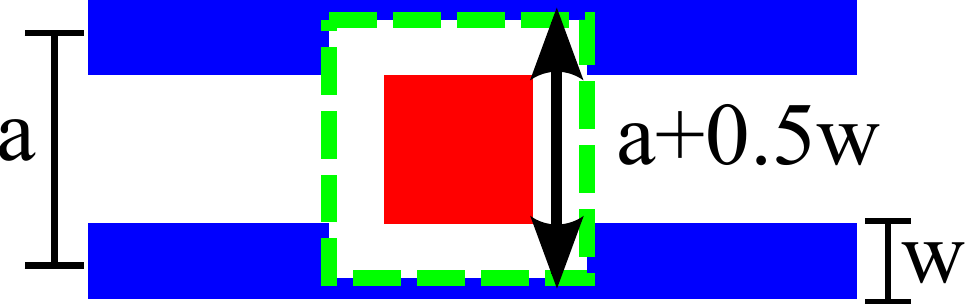}
    \caption{A2}
    \label{fig:A2-2D-illustration}
  \end{subfigure}%
  \begin{subfigure}[b]{\widthfraction\linewidth{}}
    \centering
    \includegraphics[width=.99\linewidth]{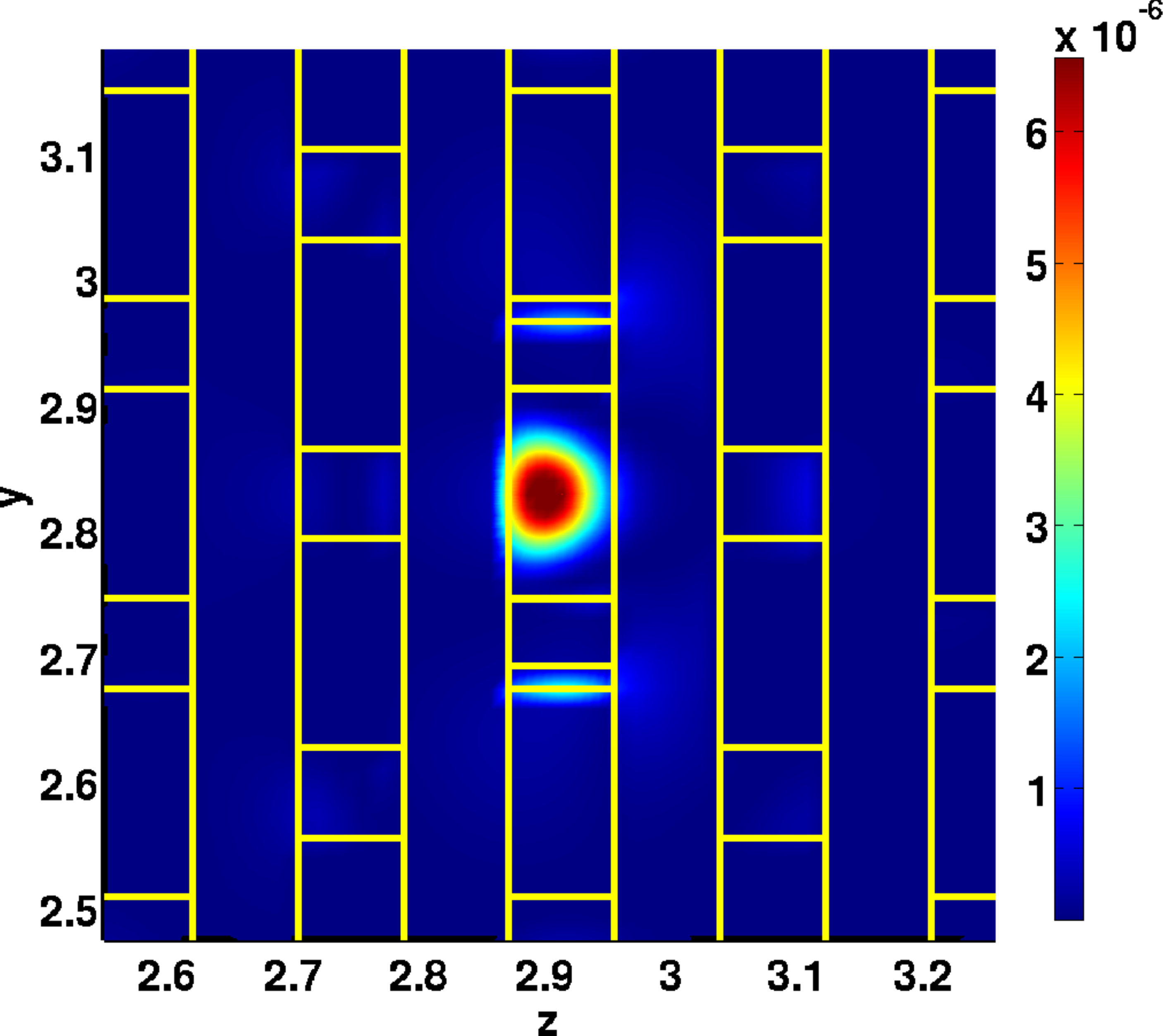}
    \caption{A2}
    \label{fig:A2-Ex-2D}
  \end{subfigure}%
  \begin{subfigure}[b]{\widthfraction\linewidth{}}
    \centering
    \includegraphics[width=.99\linewidth]{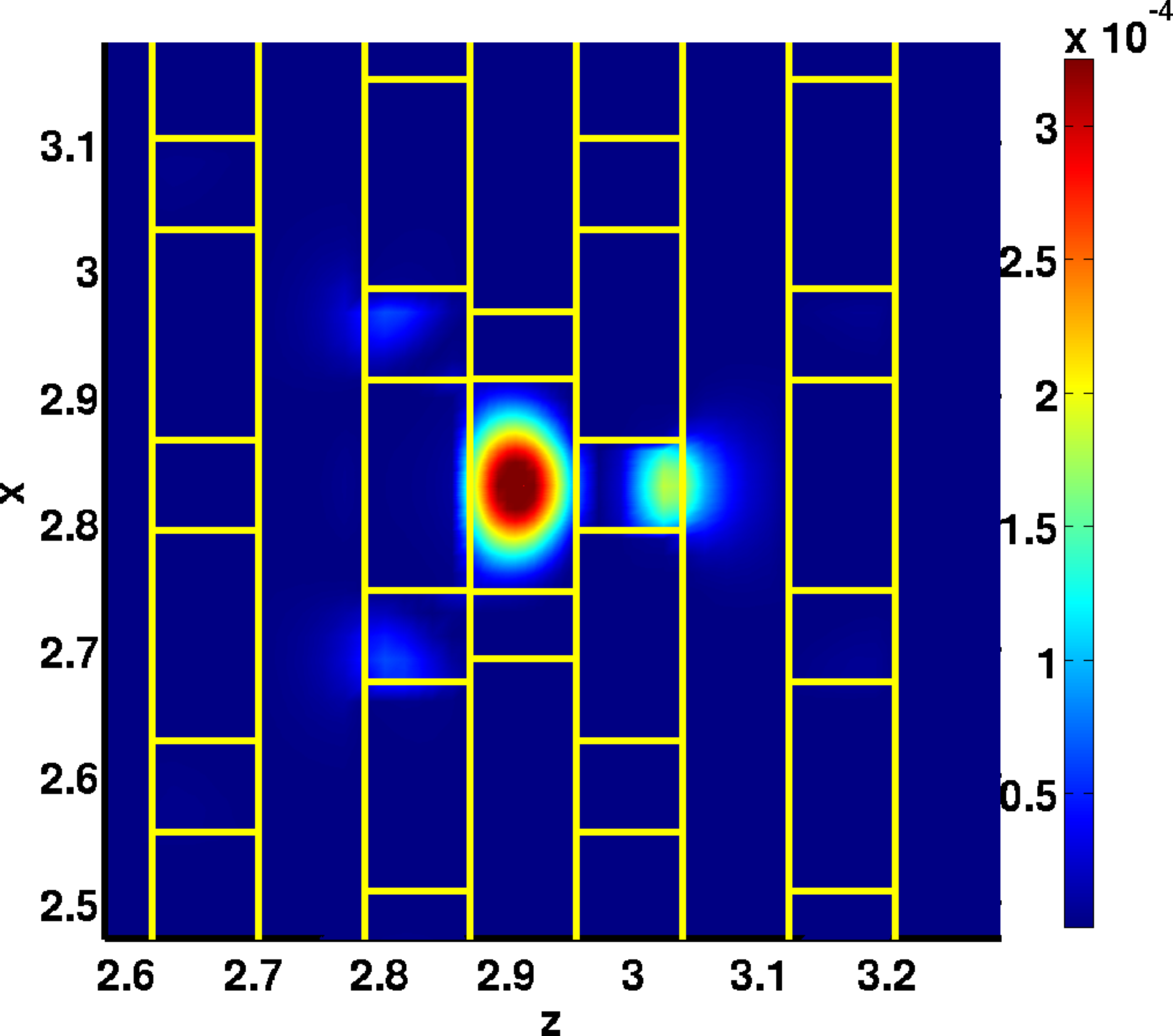}
    \caption{A2}
    \label{fig:A2-Ey-2D}
  \end{subfigure}
  \begin{subfigure}[b]{\widthfraction\linewidth{}}
    \centering
    \includegraphics[width=.99\linewidth]{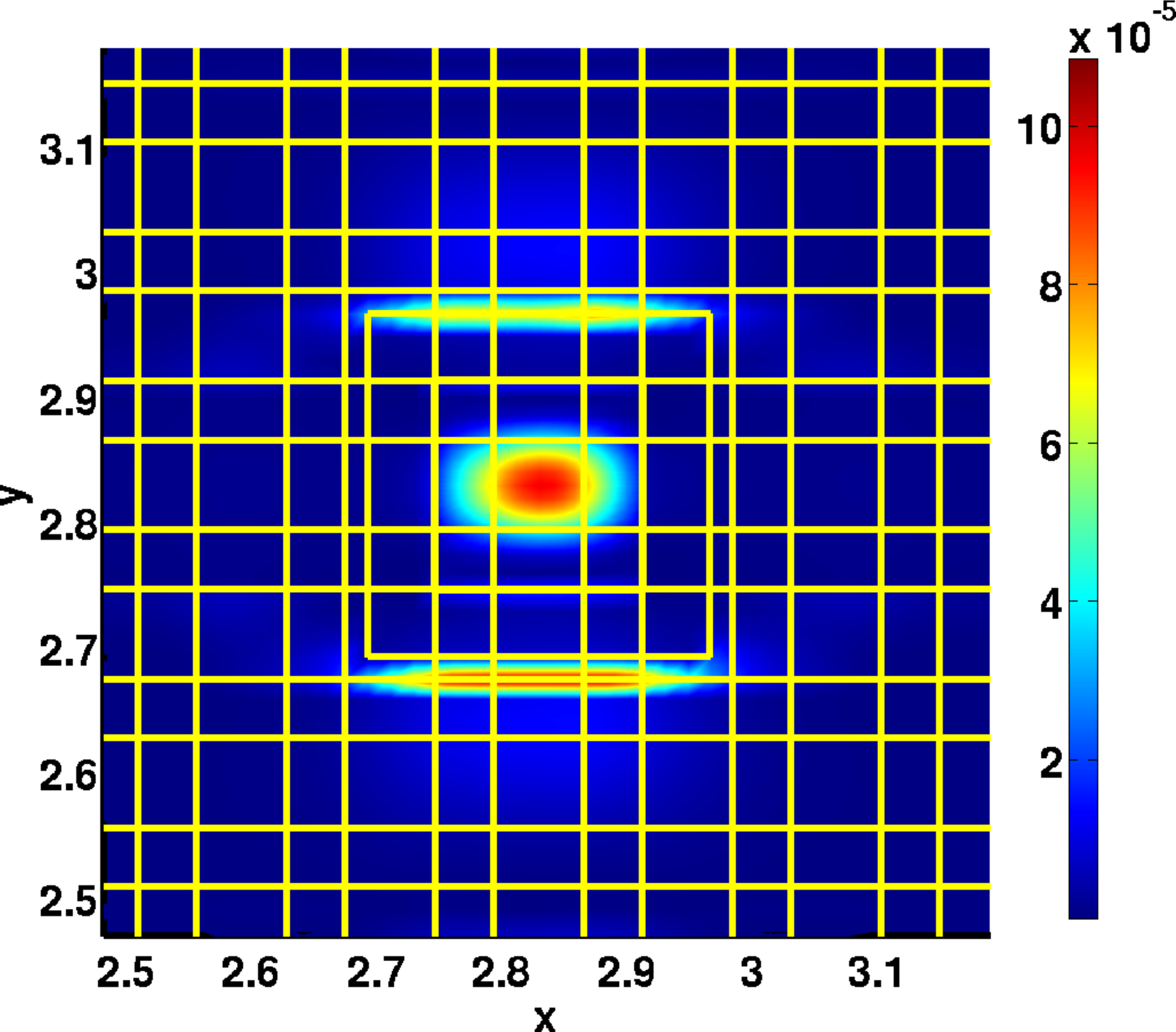}
    \caption{A2}
    \label{fig:A2-Ez-2D}
  \end{subfigure}

  \caption{
  Frequency snapshots of the energy density \ElectromagneticEnergy{} for the defects A0 (a-d), A1 (e-h) and A2 (i-l), taken in the X, Y and Z planes going through the centre of the defect.
  }
  \label{fig:figure-energySnapshots-A0-A1-A2}

\end{figure*}
}

{
\newcommand{\widthfraction}{0.24}
\newcommand{\subsize}{0.24}
\newcommand{\subsubsize}{0.99}
\newcommand{\spaceman}{0.8}

\begin{figure*}
  \centering

  \begin{subfigure}[b]{\subsize\linewidth{}}
    \centering
    \includegraphics[width=.99\linewidth]{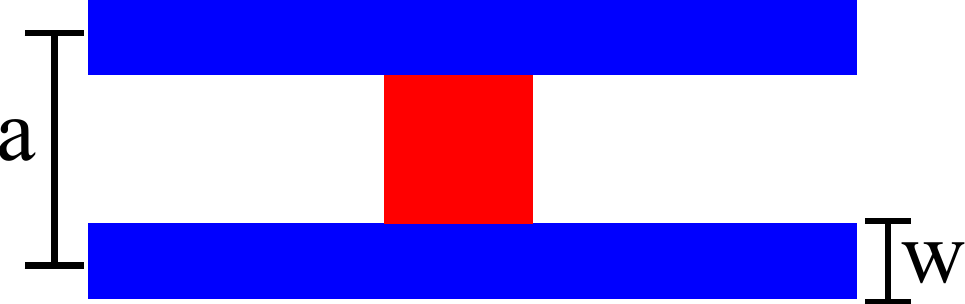}
    \caption{D1}
  \end{subfigure}%
  \begin{subfigure}[b]{\subsize\linewidth{}}
    \centering
    \includegraphics[width=.99\linewidth]{figure10a_11b_A0}
    \caption{A0}
  \end{subfigure}%
  \begin{subfigure}[b]{\subsize\linewidth{}}
    \centering
    \includegraphics[width=.99\linewidth]{figure10e_11c_A1}
    \caption{A1}
  \end{subfigure}
  \begin{subfigure}[b]{\subsize\linewidth{}}
    \centering
    \includegraphics[width=.99\linewidth]{figure10i_11d_A2}
    \caption{A2}
  \end{subfigure}

  \begin{subfigure}[b]{\widthfraction\linewidth{}}
    \centering
    \includegraphics[width=\spaceman\linewidth]{figure07d_11e}
    \caption{D1}
    \label{fig:D1-Ex-3D-v2}
  \end{subfigure}%
  \begin{subfigure}[b]{\widthfraction\linewidth{}}
    \centering
    \includegraphics[width=\spaceman\linewidth]{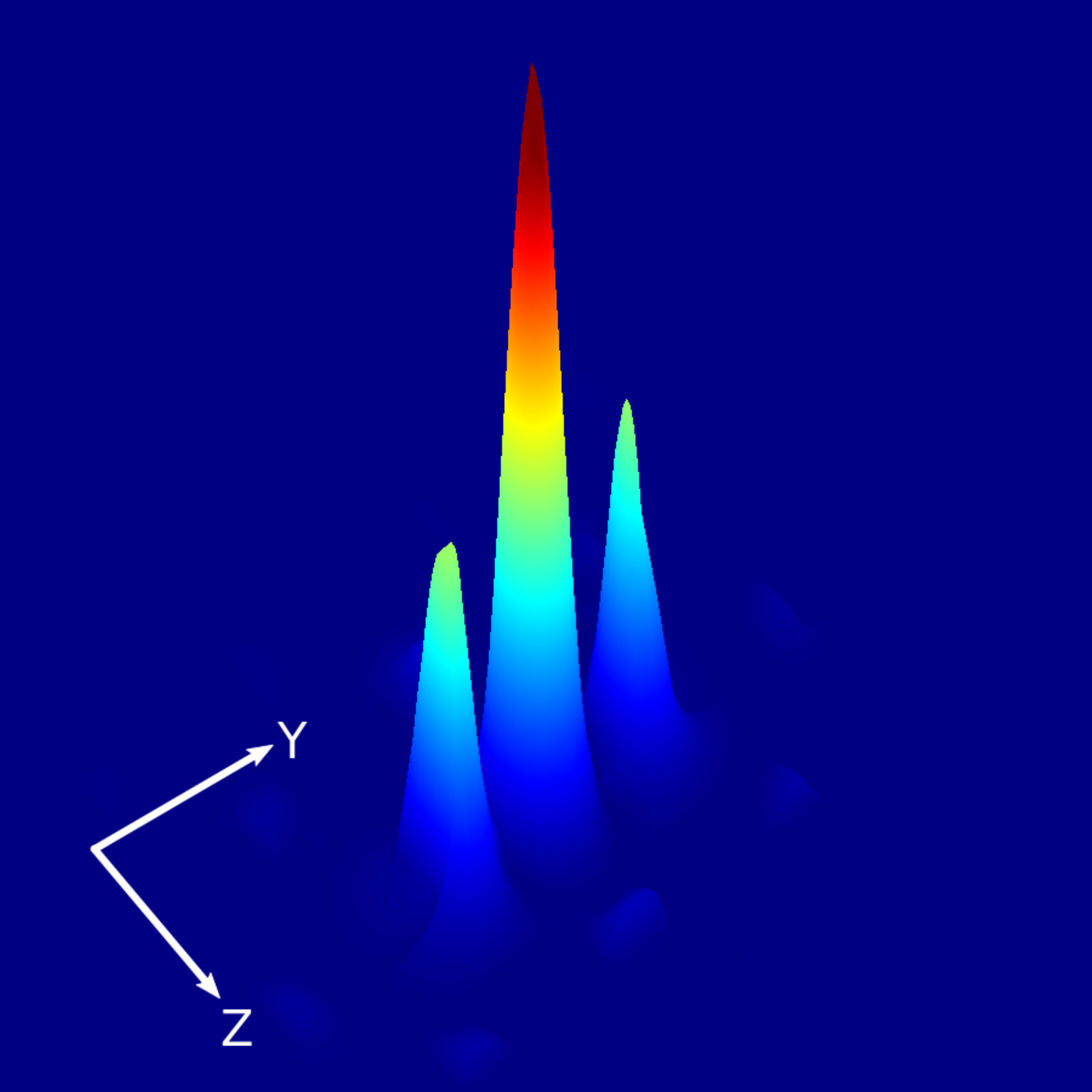}
    \caption{A0}
    \label{fig:A0-Ex-3D}
  \end{subfigure}%
  \begin{subfigure}[b]{\widthfraction\linewidth{}}
    \centering
    \includegraphics[width=\spaceman\linewidth]{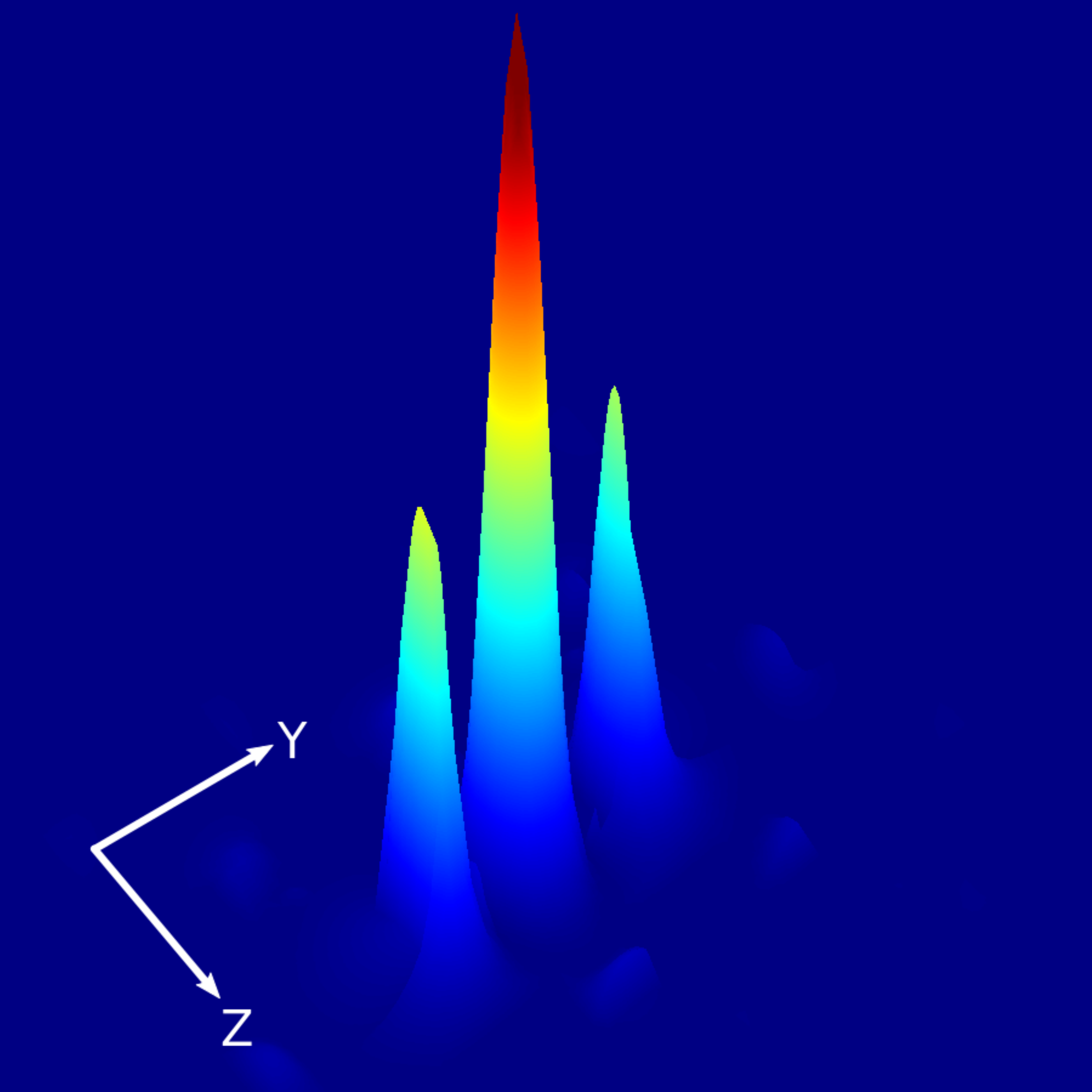}
    \caption{A1}
    \label{fig:A1-Ex-3D}
  \end{subfigure}
  \begin{subfigure}[b]{\widthfraction\linewidth{}}
    \centering
    \includegraphics[width=\spaceman\linewidth]{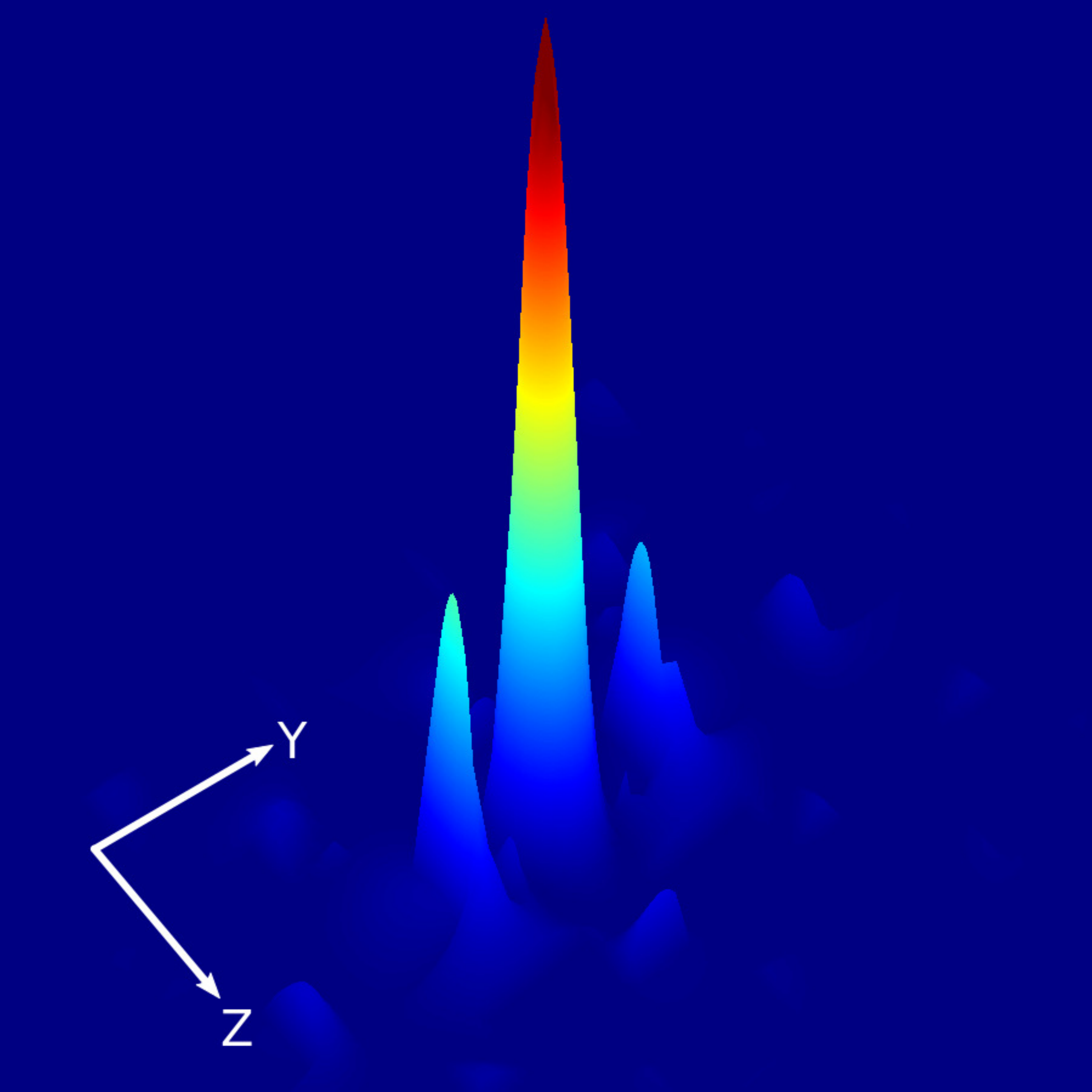}
    \caption{A2}
    \label{fig:A2-Ex-3D}
  \end{subfigure}

  \caption{
  3D view of the \EnergySnapshots{} \ElectromagneticEnergy{} for the defects D1 (a,e), A0 (b,f), A1 (c,g) and A2 (d,h). The snapshots were taken in the Y-Z planes going through the centres of the defects.
  }
  \label{fig:figure-energySnapshots-D1-A0-A1-A2}
\end{figure*}
}

\begin{figure}

  \centering
  \includegraphics[width=0.8\linewidth]{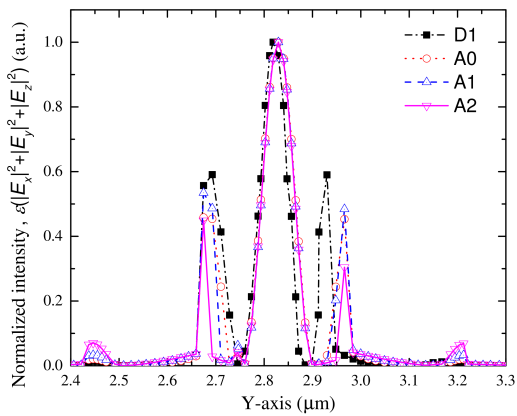}

  \caption{
  Normalized energy density \ElectromagneticEnergy{} along a line parallel to the Y axis going through the centre of the cavity for the D1 defect without air buffer and the A0, A1 and A2 defects with air buffers.
  This corresponds to a cross-section along the Y axis for $ z = 2.9134 \mu{}m $ of the \EnergySnapshots{} illustrated in figures \ref{fig:D1-Ex-2D}, \ref{fig:A0-Ex-2D}, \ref{fig:A1-Ex-2D}, \ref{fig:A2-Ex-2D} and figures \ref{fig:D1-Ex-3D-v2}, \ref{fig:A0-Ex-3D}, \ref{fig:A1-Ex-3D}, \ref{fig:A2-Ex-3D}.
  }
  \label{fig:central-cut}
\end{figure}

\begingroup
\squeezetable
  \begin{table*}
    \caption{
      \label{tab:tabular-DefectSizes}
      Summary of achievable Q-factors $Q_{uc}$, modal volumes $V_{eff}$, Purcell factors $F_p$, and leakage rates $\kappa_{uc}/(2\pi)$ at the highest-Q resonant modes ($\VerticalPeriod/\lambda$) for the defects without air buffer D0, D1 and D2.
      Simulations for the defect D0 in the $E_{x}$ direction showed no clearly distinguishable resonance modes, which is why its entries are marked \textit{n/a}.
    }
    \begin{ruledtabular}

      \begin{tabular}{llllllllll}
      Defect type  & \multicolumn{3}{c}{D0} & \multicolumn{3}{c}{D1} & \multicolumn{3}{c}{D2}\tabularnewline
      Defect size  & \multicolumn{3}{c}{$(0.25,\,0.25,\,0.5)\cdot c$} & \multicolumn{3}{c}{$(0.5,\,0.5,\,0.25)\cdot c$} & \multicolumn{3}{c}{$(0.5,\,0.5,\,0.5)\cdot c$}\tabularnewline
      Air buffer size  & \multicolumn{3}{c}{$n/a$} & \multicolumn{3}{c}{$n/a$} & \multicolumn{3}{c}{$n/a$}\tabularnewline
      $n_{def}$  & \multicolumn{3}{c}{$3.3$} & \multicolumn{3}{c}{$3.3$} & \multicolumn{3}{c}{$3.3$}\tabularnewline
      $n_{wp}$  & \multicolumn{3}{c}{$3.3$} & \multicolumn{3}{c}{$3.3$} & \multicolumn{3}{c}{$3.3$}\tabularnewline
      $n_{bf}$  & \multicolumn{3}{c}{$1$} & \multicolumn{3}{c}{$1$} & \multicolumn{3}{c}{$1$}\tabularnewline
      $c(\mu m)$  & \multicolumn{3}{c}{$0.3358$} & \multicolumn{3}{c}{$0.3358$} & \multicolumn{3}{c}{$0.3358$}\tabularnewline
      \hline
      Dipole orientation  & $E_{x}$  & $E_{y}$  & $E_{z}$  & $E_{x}$  & $E_{y}$  & $E_{z}$  & $E_{x}$  & $E_{y}$  & $E_{z}$\tabularnewline
      $c/\lambda_{0}$  & $n/a$  & $0.5890$  & $0.5530$  & $0.5255$  & $0.5058$  & $0.5300$  & $0.4938$  & $0.4922$  & $0.5007$\tabularnewline
      $\lambda_{0}(nm)$  & $n/a$  & $570.08$  & $607.23$  & $638.98$  & $663.88$  & $633.53$  & $679.99$  & $682.22$  & $670.71$\tabularnewline
      $Q_{uc}$  & $n/a$  & $3.71\times10^{2}$  & $3.34\times10^{4}$  & $7.54\times10^{5}$  & $2.73\times10^{5}$  & $8.35\times10^{4}$  & $1.35\times10^{5}$  & $1.06\times10^{5}$  & $8.05\times10^{4}$\tabularnewline
      $V_{eff}(\mu m^{3})$  & $n/a$  & $5.09\times10^{-3}$  & $1.19\times10^{-3}$  & $1.17\times10^{-3}$  & $2.44\times10^{-3}$  & $1.37\times10^{-3}$  & $2.89\times10^{-3}$  & $3.48\times10^{-3}$  & $4.56\times10^{-3}$\tabularnewline
      $V_{n}=\frac{V_{eff}}{(\lambda_{0}/n_{def})^{3}}$  & $n/a$  & $0.987$  & $0.190$  & $0.161$  & $0.299$  & $0.193$  & $0.330$  & $0.394$  & $0.543$\tabularnewline
      $F_{p}$  & $n/a$  & $2.85\times10^{1}$  & $1.33\times10^{4}$  & $3.56\times10^{5}$  & $6.92\times10^{4}$  & $3.29\times10^{4}$  & $3.10\times10^{4}$  & $2.05\times10^{4}$  & $1.13\times10^{4}$\tabularnewline
      \KappaExprTwoLine{}  & $n/a$  & $1418.48$  & $14.79$  & $0.62$  & $1.66$  & $5.66$  & $3.27$  & $4.13$  & $5.56$\tabularnewline
      $\tau_{uc}=2\pi/\kappa_{uc}$(ns)  & $n/a$  & $7.05\times10^{-4}$  & $0.07$  & $1.61$  & $0.60$  & $0.18$  & $0.31$  & $0.24$  & $0.18$\tabularnewline
      \hline
      Diamond NV centres  & \multicolumn{9}{c}{Diamond nanocrystals: $n_{os}\sim2.4$}\tabularnewline
      $\lambda_{os}(nm)$  & \multicolumn{9}{c}{$637$}\tabularnewline
      $V'_{eff}(\mu m^{3})=V_{n}\times\left(\frac{\lambda_{os}}{n_{def}}\right)^{3}$  & $n/a$  & $7.10\times10^{-3}$  & $1.37\times10^{-3}$  & $1.16\times10^{-3}$  & $2.15\times10^{-3}$  & $1.39\times10^{-3}$  & $2.37\times10^{-3}$  & $2.83\times10^{-3}$  & $3.91\times10^{-3}$\tabularnewline
      \KappaPrimeExprTwoLine{} & $n/a$  & $1269.47$  & $14.09$  & $0.62$  & $1.73$  & $5.63$  & $3.49$  & $4.43$  & $5.85$\tabularnewline
      $\tau'_{uc}=2\pi/\kappa_{uc}$(ns)  & $n/a$  & $7.88\times10^{-4}$  & $0.07$  & $1.60$  & $0.58$  & $0.18$  & $0.29$  & $0.23$  & $0.17$\tabularnewline
      $\gamma^{ZPL}/(2\pi)$(GHz)  & \multicolumn{9}{c}{$3.3\times10^{-3}$ ($2\pi/\gamma^{ZPL}=\tau^{ZPL}\sim300ns$)}\tabularnewline
      $d_{EG}^{ZPL}(C\cdot m)$  & \multicolumn{9}{c}{$3.57\times10^{-30}(=1.07D)$}\tabularnewline
      $f_{EG}^{ZPL}$  & \multicolumn{9}{c}{$0.025$}\tabularnewline
      $E_{sp}$(V/m)  & $n/a$  & $4.77\times10^{5}$  & $1.09\times10^{6}$  & $1.18\times10^{6}$  & $8.67\times10^{5}$  & $1.08\times10^{6}$  & $8.25\times10^{5}$  & $7.55\times10^{5}$  & $6.43\times10^{5}$\tabularnewline
      $g_{R}^{ZPL}/(2\pi)$(GHz)  & $n/a$  & $2.57$  & $5.85$  & $6.37$  & $4.67$  & $5.81$  & $4.45$  & $4.07$  & $3.46$\tabularnewline
      $\tau_{R}^{ZPL}=2\pi/g_{R}^{ZPL}$(ns)  & $n/a$  & $0.39$  & $0.17$  & $0.16$  & $0.21$  & $0.17$  & $0.22$  & $0.25$  & $0.29$\tabularnewline
      \end{tabular}
      
    \end{ruledtabular}
  \end{table*}
\endgroup

\begingroup
\squeezetable
  \begin{table*}
    \caption{
      \label{tab:tabular-AirBuffers}
      Summary of achievable Q-factors $Q_{uc}$, modal volumes $V_{eff}$, Purcell factors $F_p$, and leakage rates $\kappa_{uc}/(2\pi)$ at the highest-Q resonant modes ($\VerticalPeriod/\lambda$) for the defects with air buffer A0, A1 and A2.
    }
    \begin{ruledtabular}

      \begin{tabular}{llllllllll}
      Defect type & \multicolumn{3}{c}{A0} & \multicolumn{3}{c}{A1} & \multicolumn{3}{c}{A2}\tabularnewline
      Defect size & \multicolumn{3}{c}{$(0.5,\,0.5,\,0.25)\cdot c$} & \multicolumn{3}{c}{$(0.5,\,0.5,\,0.25)\cdot c$} & \multicolumn{3}{c}{$(0.5,\,0.5,\,0.25)\cdot c$}\tabularnewline
      Air buffer size & \multicolumn{3}{c}{$(a-0.5\cdot w,\, a-0.5\cdot w,\,0.25\cdot c)$} & \multicolumn{3}{c}{$(a,\, a,\,0.25\cdot c)$} & \multicolumn{3}{c}{$(a+0.5\cdot w,\, a+0.5\cdot w,\,0.25\cdot c)$}\tabularnewline
      $n_{def}$  & \multicolumn{3}{c}{$3.3$} & \multicolumn{3}{c}{$3.3$} & \multicolumn{3}{c}{$3.3$}\tabularnewline
      $n_{wp}$  & \multicolumn{3}{c}{$3.3$} & \multicolumn{3}{c}{$3.3$} & \multicolumn{3}{c}{$3.3$}\tabularnewline
      $n_{bf}$  & \multicolumn{3}{c}{$1$} & \multicolumn{3}{c}{$1$} & \multicolumn{3}{c}{$1$}\tabularnewline
      $c(\mu m)$  & \multicolumn{3}{c}{$0.3358$} & \multicolumn{3}{c}{$0.3358$} & \multicolumn{3}{c}{$0.3358$}\tabularnewline
      \hline
      Dipole orientation & $E_{x}$  & $E_{y}$  & $E_{z}$  & $E_{x}$  & $E_{y}$  & $E_{z}$  & $E_{x}$  & $E_{y}$  & $E_{z}$\tabularnewline
      $c/\lambda_{0}$  & $0.5409$  & $0.5303$  & $0.5377$  & $0.5513$  & $0.5391$  & $0.5449$  & $0.5715$  & $0.5440$  & $0.5593$\tabularnewline
      $\lambda_{0}$(nm)  & $620.86$  & $633.28$  & $624.46$  & $609.13$  & $622.88$  & $616.23$  & $587.59$  & $617.28$  & $600.35$\tabularnewline
      $Q_{uc}$  & $3.67\times10^{5}$  & $2.50\times10^{5}$  & $6.14\times10^{4}$  & $1.48\times10^{5}$  & $1.69\times10^{5}$  & $4.25\times10^{4}$  & $3.80\times10^{3}$  & $1.17\times10^{5}$  & $1.35\times10^{4}$\tabularnewline
      $V_{eff}(\mu m^{3})$  & $6.66\times10^{-4}$  & $2.09\times10^{-3}$  & $1.13\times10^{-3}$  & $7.76\times10^{-4}$  & $1.39\times10^{-3}$  & $1.26\times10^{-3}$  & $1.73\times10^{-3}$  & $9.88\times10^{-4}$  & $7.22\times10^{-4}$\tabularnewline
      $V_{n}=\frac{V_{eff}}{(\lambda_{0}/n_{def})^{3}}$  & $0.100$  & $0.296$  & $0.167$  & $0.123$  & $0.207$  & $0.193$  & $0.306$  & $0.151$  & $0.120$\tabularnewline
      $F_{p}$  & $2.78\times10^{5}$  & $6.41\times10^{4}$  & $2.79\times10^{4}$  & $9.14\times10^{4}$  & $6.21\times10^{4}$  & $1.67\times10^{4}$  & $9.45\times10^{2}$  & $5.92\times10^{4}$  & $8.58\times10^{3}$\tabularnewline
      \KappaExprTwoLine{} & $1.32$  & $1.90$  & $7.82$  & $3.32$  & $2.85$  & $11.45$  & $134.14$  & $4.13$  & $36.91$\tabularnewline
      $\tau_{uc}=2\pi/\kappa_{uc}$(ns)  & $0.76$  & $0.53$  & $0.13$  & $0.30$  & $0.35$  & $0.09$  & $0.01$  & $0.24$  & $0.03$\tabularnewline
      \hline
      Diamond NV centres & \multicolumn{9}{c}{Diamond nanocrystals: $n_{os}\sim2.4$}\tabularnewline
      $\lambda_{os}(nm)$  & \multicolumn{9}{c}{$637$}\tabularnewline
      $V'_{eff}(\mu m^{3})=V_{n}\times\left(\frac{\lambda_{os}}{n_{def}}\right)^{3}$  & $7.20\times10^{-4}$  & $2.13\times10^{-3}$  & $1.20\times10^{-3}$  & $8.87\times10^{-4}$  & $1.49\times10^{-3}$  & $1.39\times10^{-3}$  & $2.20\times10^{-3}$  & $1.09\times10^{-3}$  & $8.62\times10^{-4}$\tabularnewline
      \KappaPrimeExprTwoLine{} & $1.28$  & $1.88$  & $7.66$  & $3.17$  & $2.78$  & $11.08$  & $123.73$  & $4.01$  & $34.78$\tabularnewline
      $\tau'_{uc}=2\pi/\kappa_{uc}$(ns)  & $0.78$  & $0.53$  & $0.13$  & $0.32$  & $0.36$  & $0.09$  & $0.01$  & $0.25$  & $0.03$\tabularnewline
      $\gamma^{ZPL}/(2\pi)$(GHz)  & \multicolumn{9}{c}{$3.3\times10^{-3}$ ($2\pi/\gamma^{ZPL}=\tau^{ZPL}\sim300ns$)}\tabularnewline
      $d_{EG}^{ZPL}(C\cdot m)$  & \multicolumn{9}{c}{$3.57\times10^{-30}(=1.07D)$}\tabularnewline
      $f_{EG}^{ZPL}$  & \multicolumn{9}{c}{$0.025$}\tabularnewline
      $E_{sp}$(V/m)  & $1.50\times10^{6}$  & $8.71\times10^{5}$  & $1.16\times10^{6}$  & $1.35\times10^{6}$  & $1.04\times10^{6}$  & $1.08\times10^{6}$  & $8.57\times10^{5}$  & $1.22\times10^{6}$  & $1.37\times10^{6}$\tabularnewline
      $g_{R}^{ZPL}/(2\pi)$(GHz)  & $8.07$  & $4.69$  & $6.24$  & $7.27$  & $5.61$  & $5.82$  & $4.62$  & $6.57$  & $7.38$\tabularnewline
      $\tau_{R}^{ZPL}=2\pi/g_{R}^{ZPL}$(ns)  & $0.12$  & $0.21$  & $0.16$  & $0.14$  & $0.18$  & $0.17$  & $0.22$  & $0.15$  & $0.14$\tabularnewline
      \end{tabular}


    \end{ruledtabular}
  \end{table*}
\endgroup

The smallest mode volume and biggest coupling strength are obtained for the defect A0 (\fref{fig:A0}), with $V_{eff} \simeq \AaXMVnormalized $ and $g_{R}/(2\pi) = 8.07 GHz $.
The corresponding Q-factor $ Q \simeq \AaXQ $ is lower than for the same defect without air buffer where $Q \simeq \SbXQ $, which leads to a weaker coupling with $ 4 g_{R} / (\kappa+\gamma) = 25.09 $.
But since the Q-factor can be increased by increasing the number of periods of the woodpile structure, \cite{Ho2011:IEEE_JQE,Imagawa2012,Tajiri2014}, stronger coupling could still be achieved, still rendering this kind of air buffer useful.

\section{Conclusion}

In this article, we have investigated 3D FCC woodpile photonic crystals formed in high-index-contrast materials (GaP) and found a maximum photonic band gap (PBG) of $16\%$ at $(w/c)_{opt} \simeq 0.2145$ by using the plane-wave expansion method.
We also introduced cuboid-shaped defects with various sizes near the centre of the middle layer in the 3D PhCs and calculated the Q-factors and mode volumes of cavity modes using the FDTD method.
The best results were obtained for the defect D1 for which we found a Q-factor $ Q = 7.54 \times{} 10^{5} $ and a mode volume $ V_{eff} = 0.161 ( \lambda{}_{res}/n)^{3} $ for an excitation in the $E_{x}$ direction.
Moreover, we found that adding air buffers around the defect allows us to reduce the mode volumes.
The cuboid-shaped defect cavities with air buffer (A0 and A1) showed smaller mode volumes, especially for A0, with mode volumes down to $V_{eff}=0.100(\lambda_{res}/n)^3$, while still having high Q-factors.
These high-Q cavities (D1,D2,A0,A1) with small mode volumes would allow the observation of strong coupling of a single solid-state quantum system (NV-centre) with the cavity mode.

Fabricating our modelled structures would require \textless{}~$100nm$ feature sizes and will therefore be very challenging.
However, we are actively pursuing 3D fabrication of such structures using DLW and have already seen partial bandgaps in 3D photonic crystals down to 1400nm \cite{Chen2015}.
The use of smaller wavelengths \cite{Mueller2014} and new techniques like Stimulated-Emission-Depletion (STED) \cite{Fischer2015} in DLW might allow even higher writing resolutions and therefore bandgaps at smaller wavelengths.

\begin{acknowledgments}
This work was carried out using the computational facilities of the Advanced Computing Research Center, University of Bristol, Bristol, U.K.
We acknowledge financial support from the ERC advanced grant 247462 QUOWSS and the EU FP7 grant 618078 WASPS.
\end{acknowledgments}

\bibliographystyle{osajnl}
\bibliography{master.bib}

\end{document}